\batchmode
\makeatletter
\def\input@path{{"/home/jacob/Documents/Work/My Papers/2025-The ABL Rule and Weak Values/"}}
\makeatother
\documentclass[11pt,twoside,english]{article}
\usepackage[T1]{fontenc}
\usepackage[latin9]{inputenc}
\usepackage{babel}
\usepackage{amsmath}
\usepackage{amssymb}
\usepackage{geometry}
\usepackage{setspace}
\usepackage[pdftex,
 bookmarks=true,bookmarksnumbered=true,bookmarksopen=false,
 breaklinks=false,pdfborder={0 0 0},pdfborderstyle={},backref=false,colorlinks=false]
 {hyperref}
\hypersetup{
 pdfauthor={Jacob A. Barandes},
 pdfsubject={Physics},
 pdfkeywords={physics}}

\makeatletter

\let\originalleft\left
\let\originalright\right
\renewcommand{\left}{\mathopen{}\mathclose\bgroup\originalleft}
\renewcommand{\right}{\aftergroup\egroup\originalright}

\def\smalloverbrace#1{\mathop{\vbox{\m@th\ialign{##\crcr%
      \noalign{\kern3\p@}%
      \tiny\downbracefill\crcr\noalign{\kern3\p@\nointerlineskip}%
      $\hfil\displaystyle{#1}\hfil$\crcr}}}\limits}

\def\smallunderbrace#1{\mathop{\vtop{\m@th\ialign{##\crcr
   $\hfil\displaystyle{#1}\hfil$\crcr
   \noalign{\kern3\p@\nointerlineskip}%
   \tiny\upbracefill\crcr\noalign{\kern3\p@}}}}\limits}

\DeclareMathAlphabet{\mymathbb}{U}{bbold}{m}{n}

\makeatother

\begin{document}
\title{The Trouble with Weak Values}
\author{Jacob A. Barandes\thanks{Departments of Philosophy and Physics, Harvard University, Cambridge, MA 02138; jacob\_barandes@harvard.edu; ORCID: 0000-0002-3740-4418}
}
\date{\today}

\maketitle

\begin{abstract}
In quantum theory, a weak value is a complex number with a somewhat
technical definition: it is a ratio whose numerator is the matrix
element of a self-adjoint operator and whose denominator is the inner
product of a corresponding pair of state vectors. Weak values first
appeared in the research literature in a pair of papers in 1987 and
1988, and were originally defined as the results of a special kind
of experimental protocol involving non-disturbing measurements combined
with an explicit form of post-selection. In the years since, subsequent
papers on weak values have produced a number of important practical
spin-offs, including new methods for signal amplification and quantum-state
tomography. The present work is not concerned with those practical
spin-offs, but with historical and ongoing attempts to assign weak
values a transparent, single-system interpretation, as well as efforts
that invoke weak values to make a number of exotic claims about the
properties and behavior of individual quantum systems. This paper
challenges these interpretational claims by arguing that they involve
several forms of fallacious reasoning.
\end{abstract}

\begin{center}
\global\long\def\quote#1{``#1"}%
\global\long\def\apostrophe{\textrm{'}}%
\global\long\def\slot{\phantom{x}}%
\global\long\def\eval#1{\left.#1\right\vert }%
\global\long\def\keyeq#1{\boxed{#1}}%
\global\long\def\importanteq#1{\boxed{\boxed{#1}}}%
\global\long\def\given{\vert}%
\global\long\def\mapping#1#2#3{#1:#2\to#3}%
\global\long\def\composition{\circ}%
\global\long\def\set#1{\left\{  #1\right\}  }%
\global\long\def\setindexed#1#2{\left\{  #1\right\}  _{#2}}%

\global\long\def\setbuild#1#2{\left\{  \left.\!#1\,\right|\,#2\right\}  }%
\global\long\def\complem{\mathrm{c}}%

\global\long\def\union{\cup}%
\global\long\def\intersection{\cap}%
\global\long\def\cartesianprod{\times}%
\global\long\def\disjointunion{\sqcup}%

\global\long\def\isomorphic{\cong}%

\global\long\def\setsize#1{\left|#1\right|}%
\global\long\def\defeq{\equiv}%
\global\long\def\conj{\ast}%
\global\long\def\overconj#1{\overline{#1}}%
\global\long\def\re{\mathrm{Re\,}}%
\global\long\def\im{\mathrm{Im\,}}%

\global\long\def\transp{\mathrm{T}}%
\global\long\def\tr{\mathrm{tr}}%
\global\long\def\adj{\dagger}%
\global\long\def\diag#1{\mathrm{diag}\left(#1\right)}%
\global\long\def\dotprod{\cdot}%
\global\long\def\crossprod{\times}%
\global\long\def\Probability#1{\mathrm{Prob}\left(#1\right)}%
\global\long\def\Amplitude#1{\mathrm{Amp}\left(#1\right)}%
\global\long\def\cov{\mathrm{cov}}%
\global\long\def\corr{\mathrm{corr}}%

\global\long\def\absval#1{\left\vert #1\right\vert }%
\global\long\def\expectval#1{\left\langle #1\right\rangle }%
\global\long\def\op#1{\hat{#1}}%

\global\long\def\bra#1{\left\langle #1\right|}%
\global\long\def\ket#1{\left|#1\right\rangle }%
\global\long\def\braket#1#2{\left\langle \left.\!#1\right|#2\right\rangle }%

\global\long\def\parens#1{(#1)}%
\global\long\def\bigparens#1{\big(#1\big)}%
\global\long\def\Bigparens#1{\Big(#1\Big)}%
\global\long\def\biggparens#1{\bigg(#1\bigg)}%
\global\long\def\Biggparens#1{\Bigg(#1\Bigg)}%
\global\long\def\bracks#1{[#1]}%
\global\long\def\bigbracks#1{\big[#1\big]}%
\global\long\def\Bigbracks#1{\Big[#1\Big]}%
\global\long\def\biggbracks#1{\bigg[#1\bigg]}%
\global\long\def\Biggbracks#1{\Bigg[#1\Bigg]}%
\global\long\def\curlies#1{\{#1\}}%
\global\long\def\bigcurlies#1{\big\{#1\big\}}%
\global\long\def\Bigcurlies#1{\Big\{#1\Big\}}%
\global\long\def\biggcurlies#1{\bigg\{#1\bigg\}}%
\global\long\def\Biggcurlies#1{\Bigg\{#1\Bigg\}}%
\global\long\def\verts#1{\vert#1\vert}%
\global\long\def\bigverts#1{\big\vert#1\big\vert}%
\global\long\def\Bigverts#1{\Big\vert#1\Big\vert}%
\global\long\def\biggverts#1{\bigg\vert#1\bigg\vert}%
\global\long\def\Biggverts#1{\Bigg\vert#1\Bigg\vert}%
\global\long\def\Verts#1{\Vert#1\Vert}%
\global\long\def\bigVerts#1{\big\Vert#1\big\Vert}%
\global\long\def\BigVerts#1{\Big\Vert#1\Big\Vert}%
\global\long\def\biggVerts#1{\bigg\Vert#1\bigg\Vert}%
\global\long\def\BiggVerts#1{\Bigg\Vert#1\Bigg\Vert}%
\global\long\def\ket#1{\vert#1\rangle}%
\global\long\def\bigket#1{\big\vert#1\big\rangle}%
\global\long\def\Bigket#1{\Big\vert#1\Big\rangle}%
\global\long\def\biggket#1{\bigg\vert#1\bigg\rangle}%
\global\long\def\Biggket#1{\Bigg\vert#1\Bigg\rangle}%
\global\long\def\bra#1{\langle#1\vert}%
\global\long\def\bigbra#1{\big\langle#1\big\vert}%
\global\long\def\Bigbra#1{\Big\langle#1\Big\vert}%
\global\long\def\biggbra#1{\bigg\langle#1\bigg\vert}%
\global\long\def\Biggbra#1{\Bigg\langle#1\Bigg\vert}%
\global\long\def\braket#1#2{\langle#1\vert#2\rangle}%
\global\long\def\bigbraket#1#2{\big\langle#1\big\vert#2\big\rangle}%
\global\long\def\Bigbraket#1#2{\Big\langle#1\Big\vert#2\Big\rangle}%
\global\long\def\biggbraket#1#2{\bigg\langle#1\bigg\vert#2\bigg\rangle}%
\global\long\def\Biggbraket#1#2{\Bigg\langle#1\Bigg\vert#2\Bigg\rangle}%
\global\long\def\angs#1{\langle#1\rangle}%
\global\long\def\bigangs#1{\big\langle#1\big\rangle}%
\global\long\def\Bigangs#1{\Big\langle#1\Big\rangle}%
\global\long\def\biggangs#1{\bigg\langle#1\bigg\rangle}%
\global\long\def\Biggangs#1{\Bigg\langle#1\Bigg\rangle}%

\global\long\def\vec#1{\mathbf{#1}}%
\global\long\def\vecgreek#1{\boldsymbol{#1}}%
\global\long\def\idmatrix{\mymathbb{1}}%
\global\long\def\projector{P}%
\global\long\def\permutationmatrix{\Sigma}%
\global\long\def\densitymatrix{\rho}%
\global\long\def\krausmatrix{K}%
\global\long\def\stochasticmatrix{\Gamma}%
\global\long\def\lindbladmatrix{L}%
\global\long\def\dynop{\Theta}%
\global\long\def\timeevop{U}%
\global\long\def\hadamardprod{\odot}%
\global\long\def\tensorprod{\otimes}%

\global\long\def\inprod#1#2{\left\langle #1,#2\right\rangle }%
\global\long\def\normket#1{\left\Vert #1\right\Vert }%
\global\long\def\hilbspace{\mathcal{H}}%
\global\long\def\samplespace{\Omega}%
\global\long\def\configspace{\mathcal{C}}%
\global\long\def\phasespace{\mathcal{P}}%
\global\long\def\spectrum{\sigma}%
\global\long\def\restrict#1#2{\left.#1\right\vert _{#2}}%
\global\long\def\from{\leftarrow}%
\global\long\def\statemap{\omega}%
\global\long\def\degangle#1{#1^{\circ}}%
\global\long\def\trivialvector{\tilde{v}}%
\global\long\def\eqsbrace#1{\left.#1\qquad\right\}  }%
\global\long\def\operator#1{\operatorname{#1}}%
\par\end{center}

\begin{quotation}
``It is the sculptor's power, so often alluded to, of finding the
perfect form and features of a goddess, in the shapeless block of
marble; and his ability to chip off all extraneous matter, and let
the divine excellence stand forth for itself.'' \textemdash Robert
Allyn, 1858\nocite{Allyn:1858wsp}
\end{quotation}

\section{Introduction\label{sec:Introduction}}

In quantum theory, a \textquoteleft weak value\textquoteright{} has
a rather formal definition: it is a complex-valued numerical ratio
whose numerator consists of the matrix element of a self-adjoint operator
and whose denominator consists of the Hilbert-space inner product
of a corresponding pair of state vectors. Many formal constructions
in quantum theory admit clear physical interpretations. To what extent
is such an interpretation available for weak values? Could an interpretation
of this kind reveal underlying facts about quantum systems that would
otherwise be empirically inaccessible?

The origins of weak values go back to a 1987 paper by Aharonov, Albert,
Casher, and Vaidman (AACV) (1987)\nocite{AharonovAlbertCasherVaidman:1987sqe},
together with a 1988 paper by Aharonov, Albert, and Vaidman (AAV)
(1988)\nocite{AharonovAlbertVaidman:1988htroamoacotsoas12pctotb1}.
 According to the website of the \emph{Physical Review}, the AAV
paper alone has received over 2,200 citations, and there are many
thousands more papers in the research literature on weak values today.

The discovery of weak values was a landmark and deservedly celebrated
development in science. In particular, experimental methods for extracting
weak values have led to many important practical applications, including
new methods for amplifying detector signals, estimating small evolution
parameters, and quantum-state tomography (Dressel et al. 2014)\nocite{DresselMalikMiattoJordanBoyd:2014cuqwvbaa}.

The present work will not aim its critical analysis at those practical
applications. Instead, this paper will focus on interpretational claims
about weak values in the research literature, as well as on ways in
which weak values have been used to justify still other interpretational
claims about specific kinds of quantum systems. 

One sees these sorts of interpretational claims in the original AACV
and AAV papers, which introduced an ensemble-based protocol for indirectly
extracting weak values experimentally, with post-selection playing
a central role. These two papers were partly inspired by a 1964 paper
by Aharonov, Bergmann, and Lebowitz (ABL) (1964)\nocite{AharonovBergmannLebowitz:1964tsitqpom},
which studied an interesting kind of conditional-probability formula,
called the \textquoteleft ABL rule,\textquoteright{} that was defined
in terms \textquoteleft pre-selected\textquoteright{} and \textquoteleft post-selected\textquoteright{}
quantum states.\footnote{Although both the AACV and AAV papers used pre-selection and post-selection
in essential ways, only the AACV paper mentioned the ABL rule explicitly,
and then only in passing. The AACV and AAV papers therefore logically
stood on their own. In particular, to whatever extent the discovery
of weak values has been a success, that success does not provide retroactive
evidentiary support for the validity of the ABL rule (Barandes 2026)\nocite{Barandes:2026taratpops}.}

These interpretational claims have generated much controversy over
the years. A prominent example was a flurry of arguments and counterarguments
arising from a critical paper by Ferrie and Combes that was published
in \emph{Physical Review Letters} in 2014 and centered on, among
other things, questions over whether weak values were truly quantum-mechanical
(Ferrie, Combes 2014; Vaidman 2014; Cohen 2014; Aharonov, Rohrlich
2014; Brodutch 2015; Ferrie, Combes 2015; Romito, Jordan, Aharonov,
Gefen 2016)\nocite{FerrieCombes:2014htroasctctotb1h,Vaidman:2014cohtroasctctotb1h,Cohen:2014acohtroasctctotb1h,AharonovRohrlich:2014cohtroasctctotb1h,Brodutch:2015cohtroasctctotb1h,FerrieCombes:2015facr,RomitoJordanAharonovGefen:2016wvaqycboi}.

The AACV paper introduced an explicit example that the AAV paper later
developed into a more general framework. The present work will therefore
focus primarily on the AAV paper.

Section~\ref{sec:Preliminaries} will review several preliminary
topics, including relevant fallacies of reasoning and the textbook
Dirac-von Neumann axioms of quantum theory. Section~\ref{sec:Weak-Values}
will start with a first look at weak values, and then proceed to lay
out the standard experimental protocol for extracting them, followed
by a discussion of their interpretation and history. Section~\ref{sec:Questionable-Applications-of-Weak-Values}
will present a critical analysis of two foundational applications
of weak values\textemdash to ``quantum Cheshire Cats'' and to Bohmian
trajectories. Section~\ref{sec:Conclusion} will conclude with a
brief summary and a look at larger issues.

\section{Preliminaries\label{sec:Preliminaries}}

\subsection{Three Relevant Fallacies of Reasoning\label{subsec:Three-Relevant-Fallacies-of-Reasoning}}

The critical analysis ahead will make explicit and repeated reference
to three types of fallacious reasoning introduced in other work (Barandes 2026)\nocite{Barandes:2026taratpops}:
the ensemble fallacy, the post-selection fallacy, and the measurementist
fallacy. It will be worth reviewing these fallacies briefly here.

In physics, one sometimes studies a single object or system, and sometimes
one instead imagines or physically constructs a large number, or ensemble,
of systems that share some important collection of key characteristics.

An observable feature that one can operationally assign to a single
system will, appropriately, be called a single-system observable.
An example might be the mass of a star, or the spectrum of possible
measurement outcomes belonging to some observable of a quantum system.
By contrast, an observable feature that only makes sense at the irreducibly
collective or emergent level of an ensemble as a whole will be called
an ensemble observable.

Although it is sometimes convenient to study a single-system observable
by making use of an ensemble, such a method does not collapse the
categorical distinction between single-system observables and ensemble
observables (Hance, Rarity, Ladyman 2023, Section VI)\nocite{HanceRarityLadyman:2023wvatpoaqp}.
Collapsing that categorical distinction constitutes a form of fallacious
reasoning: 

\begin{equation}
\left.\begin{minipage}{\columnwidth}
\leftskip=10pt 
\rightskip=60pt 

The Ensemble Fallacy: The category error of attempting to identify
a single-system observable with an ensemble observable, or to draw
direct inferences about a single-system observable from an ensemble
observable without independent rigorous justification.\label{eq:DefEnsembleFallacy}

\end{minipage}\hspace{-50pt}\right\}
\end{equation}

In statistics, one needs to be very careful to ensure that the way
in which one implicitly or explicitly culls an ensemble down to a
subensemble does not introduce spurious correlations and other statistical
artifacts, called selection biases (Hernán, Hernández-\'{D}\i az,
Robins 2004)\nocite{HernanHernandez-DiazRobins:2004asatsb}. A famous
example is Berkson's paradox (Berkson 1947)\nocite{Berkson:1946lotaoftathd},
or collider bias (Cole et al. 2009)\nocite{ColePlattSchistermanChuWestreichRichardsonPoole:2009ibdtcoac},
which refers to problems that arise if one is insufficiently careful
about conditioning an ensemble by a form of after-the-fact selection,
or post-selection. Berkson, in particular, noticed that when looking
at patients in a hospital, certain diseases were negatively correlated,
but this negative correlation was a form of selection bias, because
if a patient did not have the symptoms of one disease, then the patient
would need to have symptoms of a different disease in order to be
admitted to the hospital in the first place.

Yet, rather than trying to avoid or control for accidental post-selection,
some papers in the quantum-foundations research literature actually
\emph{embrace} post-selection, and then ascribe their exotic-looking
statistical results to the mysterious nature of quantum theory itself.
However, the mere fact that one is working with quantum systems does
not give one license to make the sorts of inferences in the midst
of post-selection that would otherwise be suspect in conventional
statistics. It is worth giving this sort of widespread error a name:

\begin{equation}
\left.\begin{minipage}{\columnwidth}
\leftskip=10pt 
\rightskip=60pt 

The Post-Selection Fallacy: Drawing erroneous conclusions about a
system or ensemble due to the implicit or explicit use of post-selection.\label{eq:DefPostSelectionFallacy}

\end{minipage}\hspace{-50pt}\right\}
\end{equation}

When all else fails, it can be tempting to declare that a favored
interpretation of a theoretical quantity holds, or that work aimed
at developing that theoretical quantity is worthwhile, merely because
that theoretical quantity can be measured experimentally in the laboratory.
This view is itself a fallacious form of reasoning, and merits having
its own name: 

\begin{equation}
\left.\begin{minipage}{\columnwidth}
\leftskip=10pt 
\rightskip=60pt 

The Measurementist Fallacy (or Measurementism): If a quantity of ambiguous
interpretation can be measured experimentally, then experiments alone
can provide confirmatory support for a favored interpretation of that
quantity, or a justification for theoretical work that led to the
consideration of that quantity.\label{eq:DefMeasurementistFallacy}

\end{minipage}\hspace{-50pt}\right\}
\end{equation}

\subsection{The Dirac-von Neumann Axioms\label{subsec:The-Dirac-von-Neumann-Axioms}}

To establish the ground rules for the discussion ahead, it will be
prudent to lay out the Dirac-von Neumann (DvN) axioms (Dirac 1930,
von Neumann 1932)\nocite{Dirac:1930pofm,vonNeumann:1932mgdq}, which
underlie orthodox or textbook quantum theory:
\begin{enumerate}
\item Quantum states: Each quantum system is assigned a Hilbert space of
some finite or infinite dimension. In the most general case, the system's
quantum state is a self-adjoint, positive semidefinite operator with
unit trace, called a density matrix, or density operator, $\densitymatrix$:
\begin{equation}
\densitymatrix=\densitymatrix^{\adj}\geq0,\quad\operator{tr}\densitymatrix=1.\label{eq:DvNDensityMatrix}
\end{equation}
 If the system's density matrix has rank greater than one, then it
is said to be mixed. In the special case in which the system's density
matrix is rank-one, or pure, it can be factorized as the outer product
of a unit-norm vector $\ket{\Psi}$ with its adjoint $\bra{\Psi}\defeq\ket{\Psi}^{\adj}$:
\begin{equation}
\densitymatrix=\ket{\Psi}\bra{\Psi},\quad\braket{\Psi}{\Psi}=1\quad\left[\textrm{if }\rho\textrm{ is rank-one}\right].\label{eq:DvNDensityMatrixRankOneFactorizeStateVector}
\end{equation}
 This vector, called a state vector, or wave function, is uniquely
defined up to an irrelevant overall phase factor: 
\begin{equation}
\ket{\Psi}\cong e^{i\theta}\ket{\Psi}.\label{eq:DvNStateVectorPhaseInv}
\end{equation}
 Given multiple quantum systems that are subsystems of a composite
system, their individual Hilbert spaces $\hilbspace_{1},\dots,\hilbspace_{n}$
combine as a tensor product to define the Hilbert space of the composite
system: 
\begin{equation}
\hilbspace_{\textrm{composite}}=\hilbspace_{1}\tensorprod\cdots\tensorprod\hilbspace_{n}.\label{eq:DvNCompositeHilbertSpaceTensorProduct}
\end{equation}
 The density matrices of the subsystems are related to the density
matrix of the composite system via the partial-trace operation: 
\begin{equation}
\densitymatrix_{1}=\operator{tr}_{\textrm{not }1}\left[\densitymatrix\right],\quad\dots,\quad\densitymatrix_{n}=\operator{tr}_{\textrm{not }n}\left[\densitymatrix\right].\label{eq:DvNPartialTraceDensityMatrices}
\end{equation}
\item Unitary time evolution: The quantum state of a closed quantum system,
meaning a quantum system that is not engaged in mutual interactions
or information exchange with any other quantum systems, changes with
time according to a time-indexed family of unitary transformations
that make up the system's time-evolution operator $\timeevop\left(t\from t_{0}\right)$:
\begin{equation}
\densitymatrix\left(t\right)=\timeevop\left(t\from t_{0}\right)\densitymatrix\left(t_{0}\right)\timeevop^{\adj}\left(t\from t_{0}\right),\quad\ket{\Psi\left(t\right)}=\timeevop\left(t\from t_{0}\right)\ket{\Psi\left(t_{0}\right)}.\label{eq:DvNTimeEvOpOnDensityMatrixStateVector}
\end{equation}
 If the time-evolution operator is sufficiently smooth as a function
of time, then one can define the Hamiltonian as the self-adjoint operator
\begin{equation}
H\left(t\right)\defeq i\hbar\frac{\partial\timeevop\left(t\from t_{0}\right)}{\partial t}\timeevop^{\adj}\left(t\from t_{0}\right)=H^{\adj}\left(t\right),\label{eq:DvNDefHamiltonian}
\end{equation}
 where $\hbar$ is the reduced Planck constant. One can then express
the time evolution of the system's quantum state as a first-order
differential equation\textemdash the von Neumann equation for density
matrices or the Schrödinger equation for state vectors: 
\begin{equation}
i\hbar\frac{\partial\densitymatrix\left(t\right)}{\partial t}=\left[H\left(t\right),\densitymatrix\left(t\right)\right],\quad i\hbar\frac{\partial\ket{\Psi\left(t\right)}}{\partial t}=H\left(t\right)\ket{\Psi\left(t\right)}.\label{eq:DvNVonNeumannEqSchroEq}
\end{equation}
\item Observables: A single quantum system has an associated collection
of observables, each of which is represented by a self-adjoint operator:
\begin{equation}
A=A^{\adj}.\label{eq:DvNObservableSelfAdjoint}
\end{equation}
  If a measurement is carried out on an observable $A$, then the
possible numerical outcomes are given by the operator's eigenvalue
spectrum $\spectrum\left(A\right)$.
\item The Born rule: Given a self-adjoint operator $A$ representing one
of the quantum system's observables, let $\projector_{a}$ denote
a self-adjoint, idempotent projection operator or eigenprojector associated
with an eigenvalue $a$ in the spectrum $\spectrum\left(A\right)$:\footnote{For mathematical simplicity, the spectrum $\sigma\left(A\right)$
will be assumed here to be countable.} 
\begin{equation}
\projector_{a}=\projector_{a}^{\adj}=\projector_{a}^{2},\quad A\projector_{a}=a\projector_{a}.\label{eq:DvNProjectors}
\end{equation}
 These eigenprojectors form a projection-valued measure, or PVM, meaning
that they satisfy the conditions of mutual exclusivity, 
\begin{equation}
\projector_{a}\projector_{a^{\prime}}=\delta_{aa^{\prime}}\projector_{a},\label{eq:DvNProjectorsMutuallyExclusive}
\end{equation}
 where $\delta_{aa^{\prime}}$ is the usual Kronecker delta, 
\begin{equation}
\delta_{aa^{\prime}}=\begin{cases}
1 & \textrm{for }a=a^{\prime},\\
0 & \textrm{for }a\ne a^{\prime},
\end{cases}\label{eq:DvNKroneckerDelta}
\end{equation}
 as well as a completeness relation, or resolution of the identity,
\begin{equation}
\sum_{a}\projector_{a}=\idmatrix,\label{eq:DvNCompletenessRelation}
\end{equation}
 where $\idmatrix$ is the identity operator on the quantum system's
Hilbert space $\hilbspace$. Then the probability $p\left(a\right)$
with which a measurement of $A$ will yield the eigenvalue $a$ is
given by the Born rule: 
\begin{equation}
p\left(a\right)=\operator{tr}\left(\projector_{a}\densitymatrix\right).\label{eq:DvNBornRule}
\end{equation}
 If the system's density matrix is rank-one, so that one can instead
work with a state vector $\ket{\Psi}$, then the Born rule takes the
form 
\begin{equation}
p\left(a\right)=\operator{tr}\left(\projector_{a}\ket{\Psi}\bra{\Psi}\right)=\bra{\Psi}\projector_{a}\ket{\Psi}.\label{eq:DvNBornRuleStateVector}
\end{equation}
 If the eigenprojector $\projector_{a}$ is likewise rank-one, so
that it admits a factorization of the form 
\begin{equation}
\projector_{a}=\ket a\bra a,\label{eq:DvNProjectorFactorizable}
\end{equation}
 where $\ket a$ is an eigenvector of $A$ with eigenvalue $a$, 
\begin{equation}
A\ket a=a\ket a,\label{eq:DvNEigenvalueEq}
\end{equation}
 then the Born rule takes the simpler form 
\begin{equation}
p\left(a\right)=\verts{\braket a{\Psi}}^{2}.\label{eq:DvNBornRuleSimplest}
\end{equation}
  Given the Born rule, it follows that the expectation value $\expectval A$
of an observable, meaning the statistical average of its possible
numerical outcomes $a$ weighted by their corresponding measurement-outcome
probabilities $p\left(a\right)$, 
\begin{equation}
\expectval A\defeq\sum_{a}a\,p\left(a\right),\label{eq:DvNDefExpectationValue}
\end{equation}
 can be calculated from a density matrix $\densitymatrix$ according
to 
\begin{equation}
\expectval A=\operator{tr}\left(A\densitymatrix\right),\label{eq:DvNExpectationValueFromTrace}
\end{equation}
 and can be calculated from a state vector $\ket{\Psi}$ according
to 
\begin{equation}
\expectval A=\bra{\Psi}A\ket{\Psi}.\label{eq:DvNExpectationValueFromInnerProduct}
\end{equation}
\item Collapse: At the end of a measurement that yields an eigenvalue $a$
corresponding to an eigenprojector $\projector_{a}$ of an observable
$A$, the system's quantum state reduces or collapses via a projection
by $\projector_{a}$, followed by a renormalization to restore the
trace or norm back to $1$: 
\begin{equation}
\densitymatrix\mapsto\frac{\projector_{a}\densitymatrix\projector_{a}}{\tr\left(\projector_{a}\densitymatrix\right)},\qquad\ket{\Psi}\mapsto\frac{\projector_{a}\ket{\Psi}}{\sqrt{\bra{\Psi}\projector_{a}\ket{\Psi}}}.\label{eq:DvNCollapse}
\end{equation}
 
\end{enumerate}
One can generalize these axioms to accommodate positive-operator-valued
measures, or POVMs, but that generalization will not be needed in
this paper.

According to the DvN axioms, only \emph{measurements} are capable
of triggering the third, fourth, and fifth axioms above, and only
the fifth axiom takes the system to a final quantum state that \emph{singles
out} a specific measurement outcome. However, the DvN axioms famously
do not define what a measurement is, or what kinds of systems can
carry out measurements, so the DvN axioms are manifestly incomplete.
This specific manifestation of incompleteness is called the measurement
problem.\footnote{Claims arise from time to time that decoherence alone is capable of
solving the measurement problem. By itself, however, decoherence does
not have have the resources to single out a unique measurement outcome,
let alone with a specific probability, so further interpretational
steps that go beyond the DvN axioms are necessary (Bacciagaluppi 2025)\nocite{Bacciagaluppi:2025trodiqm}.
Due to space limitations, this paper will not say anything more about
debates over the measurement problem.}

A conceptually distinct issue is that even if one had a rigorous definition
of which processes counted as measurements and which did not, there
would still ostensibly be a gap between the narrow category of measurement
processes and the seemingly much larger category of non-measurement
phenomena that we think are happening around us all the time. This
discrepancy is called the category problem in other work (Barandes 2025)\nocite{Barandes:2025tsqc}.

One cannot brush aside the category problem merely by appealing to
the probabilities and expectation values provided by the DvN axioms,\footnote{One finds just such an appeal in a popular textbook on quantum mechanics
by Shankar (1994, Chapter 6, ``The Classical Limit'')\nocite{Shankar:1994pqm}.} because those are \emph{measurement-outcome} probabilities and \emph{measurement-outcome}
expectation values, where the latter are numerical measurement outcomes
statistically weighted by measurement-outcome probabilities. These
are not probabilities or expectation values of phenomena just \emph{happening}
or \emph{existing} in a broader categorical sense. Moreover, unless
one is willing to engage in serious foundational work, one cannot
 choose to re-interpret all the measurement-outcome probabilities
of the DvN axioms as referring to phenomena happening or existing,
due to several no-go theorems, including the Kochen-Specker theorem
(Bell 1966; Kochen, Specker 1967)\nocite{Bell:1966otpohviqm,KochenSpecker:1967phvqm}.

This paper's use of orthodox or textbook quantum theory, based on
the DvN axioms, should not be taken as an endorsement of the textbook
theory or the DvN axioms. The measurement problem and the category
problem make clear that the textbook theory is likely incomplete,
and there are several candidate reformulations or \textquoteleft interpretations\textquoteright{}
of quantum theory that seek to replace the DvN axioms with a more
internally consistent set of axioms. However, these alternative reformulations
generally make the same empirical predictions as the textbook theory
for the sorts of experimental protocols that are related to weak values,
so the textbook theory will be sufficient for the purposes of this
paper.

Of course, there may be some new axiomatic framework for quantum theory
on which the critical analysis presented by this paper no longer succeeds.
It would be exciting to see the development of such a framework. This
paper's overall argument is that unless or until such a new framework
is found, it is not possible to sustain certain widely held interpretative
claims about weak values.

It will be important to remember that the DvN axioms assign observables
and their possible measurement outcomes at the level of a \emph{single}
quantum system. Given a single quantum system and a particular observable,
one is justified in talking about which outcomes a measurement can
yield for that single system, collectively forming the spectrum of
that observable. Immediately after obtaining one such measurement
outcome, one then applies the collapse axiom to that single system,
thereby locking in that measurement outcome.

Importantly, the collapse axiom ensures that if the same observable
is quickly measured again, by \emph{any other} measuring device or
observer, then the measurement outcome will be the same as before,
within small error bars. This key consequence of the collapse axiom
strongly motivates associating the given observable with that single
system, at least in an empirical and operational sense.

In addition, notice that the measurement-outcome probabilities and
expectation values provided by the DvN axioms are fully defined once
one specifies a single system's quantum state together with the relevant
self-adjoint operators belonging to that single system. Furthermore,
these probabilities and expectation values \emph{refer} to single-system
measurement outcomes. However, the DvN probabilities and expectation
values \emph{themselves} can only be measured by considering large
ensembles of systems. Hence, these probabilities and expectation values
\emph{themselves} are not single-system observables, in the strict
empirical or operational sense used in this paper, and to treat them
otherwise would be to commit the ensemble fallacy \eqref{eq:DefEnsembleFallacy}.\footnote{The AAV paper argues that in the limit in which a measuring device's
pointer variable has a very large variance, the resulting measurement-outcome
probability distribution for the observable to be measured will be
very slightly peaked at the expectation value of that observable.
However, this expectation value is then a property of a probability
distribution, and, further, arises emergently from the interaction
of the measuring device with the subject system as a unified whole,
rather than being a single-system observable of the subject system
alone. More importantly, as the AAV paper immediately points out,
``One measurement like this will give no information,'' and then
explains that this expectation value can only be measured for empirical
and operational purposes by using a physical ensemble, so it is manifestly
an ensemble observable.}

In particular, if one encodes some sort of pattern into the probability
of a given system, then that pattern will not be a single-system observable,
but an ensemble observable. This observation will turn out to be important
for the case of weak values, which are an abstract generalization
of expectation values and not a generalization of single-system observables.\footnote{In the philosophy of probability, there are views, such as subjective
or propensity theories (Hájek 2023)\nocite{Hajek:2023iop}, on which
one can try to assign probabilities a metaphysical meaning as \emph{properties}
attached to a single system. However, empirically or operationally
speaking, one can only measure probabilities at the level of an ensemble,
so according to this paper's notion of an ensemble observable, probabilities
are ensemble observables.}

\section{Weak Values\label{sec:Weak-Values}}

\subsection{A First Look at Weak Values\label{subsec:A-First-Look-at-Weak-Values}}

Given a quantum system with an observable represented by a self-adjoint
operator $A=A^{\adj}$ on its Hilbert space $\hilbspace$, together
with a pair of state vectors $\ket{\Psi},\ket{\Psi^{\prime}}$ in
its Hilbert space that have nonzero inner product $\braket{\Psi^{\prime}}{\Psi}\ne0$,
the corresponding \textquoteleft weak value\textquoteright{} $A_{w}$
is formally defined as the complex-valued ratio 
\begin{equation}
A_{w}\defeq A_{w}\left(\Psi^{\prime},\Psi\right)\defeq\frac{\bra{\Psi^{\prime}}A\ket{\Psi}}{\braket{\Psi^{\prime}}{\Psi}}\quad\left[\textrm{assuming }\braket{\Psi^{\prime}}{\Psi}\ne0\right].\label{eq:DefWeakValueGeneral}
\end{equation}
 If $\ket{\Psi^{\prime}}=\ket{\Psi}$ (up to an irrelevant phase factor),
then the unit-norm condition $\braket{\Psi}{\Psi}=1$ from \eqref{eq:DvNDensityMatrixRankOneFactorizeStateVector},
together with the usual formula for the expectation value $\expectval A=\bra{\Psi}A\ket{\Psi}$
of an observable from \eqref{eq:DvNExpectationValueFromInnerProduct},
implies that the weak value reduces to the real-valued expectation
value, 
\begin{equation}
A_{w}\left(\Psi,\Psi\right)=\expectval A,\label{eq:WeakValueSymmetricAsExpectationValue}
\end{equation}
 as noted by Vaidman (1996a)\nocite{Vaidman:1996wmeor}.\footnote{Indeed, some references write weak values \emph{in general} using
the bracket notation $\expectval A_{w}$ (for example, Wiseman 2007;
Aharonov, Popescu, Rohrlich, Skrzypczyk 2013; Cohen, Pollak 2018)\nocite{Wiseman:2007gbmiwvab,AharonovPopescuRohrlichSkrzypczyk:2013qcc,CohenPollak:2018dowvoqouosm}.} In the even more special case in which $\ket{\Psi^{\prime}}=\ket{\Psi}=\ket a$
is an eigenvector of $A$ (again up to irrelevant phase factors),
the weak value reduces to the numerical measurement result $a$, 
\begin{equation}
A_{w}\left(a,a\right)=a,\label{eq:WeakValueReducesToMeasurementOutcome}
\end{equation}
 which is again real-valued.

There is nothing about the definition \eqref{eq:DefWeakValueGeneral}
of weak values, or about the special cases \eqref{eq:WeakValueSymmetricAsExpectationValue}
and \eqref{eq:WeakValueReducesToMeasurementOutcome}, that would immediately
explain why they are called \textquoteleft weak values.\textquoteright{}
The reason for the \textquoteleft weak\textquoteright{} modifier in
their name is that the standard experimental protocol for extracting
weak values requires special kinds of interactions between subject
systems and measuring devices. These special kinds of interactions
are called \textquoteleft weak measurements\textquoteright{} because
they only slightly perturb the systems that are involved. Weak measurements
will be reviewed in Subsection~\ref{subsec:The-AAV-Experimental-Protocol},
and were originally introduced in both the 1987 AACV paper (Aharonov,
Albert, Casher, Vaidman 1987)\nocite{AharonovAlbertCasherVaidman:1987sqe}
and in the 1988 AAV paper (Aharonov, Albert, Vaidman 1988)\nocite{AharonovAlbertVaidman:1988htroamoacotsoas12pctotb1}.

Outside of the special case $\ket{\Psi^{\prime}}=\ket{\Psi}$, the
DvN axioms reviewed in Subsection~\ref{subsec:The-Dirac-von-Neumann-Axioms}
do not assign a weak value $A_{w}$ any obvious meaning or interpretation.
A glance at the definition \eqref{eq:DefWeakValueGeneral} does not
inspire confidence. A generic weak value is typically complex-valued,
and is not a single-system observable, such as the numerical outcome
in the spectrum of any observable.\footnote{In a 2014 reply to the critical analysis of Ferrie and Combes (2014)\nocite{FerrieCombes:2014htroasctctotb1h},
Vaidman wrote: ``{[}A w{]}eak value of a variable $A$ is a property
of a single quantum system pre-selected in a state $\ket{\psi}$ and
post-selected in a state $\ket{\phi}$'' (Vaidman 2014)\nocite{Vaidman:2014cohtroasctctotb1h},
but post-selection is manifestly an ensemble concept and not a single-system
concept.} Also, a weak value depends on a \emph{pair} of distinct state vectors
that are not related to each other by the sort of unitary time evolution
characteristic of a closed system. Moreover, a weak value only exists
contingently, because it is undefined if the two state vectors in
its definition have vanishing inner product. Nor is a generic weak
value a probability or an expectation value, both of which are ensemble
observables that \emph{refer} to single-system observables.

From its definition \eqref{eq:DefWeakValueGeneral}, a generic weak
value is quite clearly an ensemble observable that does not refer
to a single-system observable, at least as far as the DvN axioms are
concerned. There might well be some new axiomatic foundation for quantum
theory on which a generic weak value could be a single-system observable,
or at least an ensemble observable that \emph{refers} to a single-system
observable, but any such interpretation of a weak value would appear
to require constructing a new such axiomatic foundation.

Standard experimental methodologies, to be reviewed in Subsection~\ref{subsec:The-AAV-Experimental-Protocol},
extract weak values in only an approximate way, and depend crucially
on the suspect use of post-selection, which is arguably not an innocent
or innocuous practice (Barandes 2026)\nocite{Barandes:2026taratpops}.
For present purposes, it will be useful to present a less practical
but exact operational procedure for obtaining a weak value that does
not involve post-selection.

Given a quantum system, let $A$ be a self-adjoint operator representing
one of the system's observables, let $\ket{\Psi}$ be any unit-norm
vector in the system's Hilbert space, and define a corresponding self-adjoint,
idempotent projection operator 
\begin{equation}
\projector_{\Psi}\defeq\ket{\Psi}\bra{\Psi}=\projector_{\Psi}^{\adj}=\projector_{\Psi}^{2},\qquad\operator{tr}\projector_{\Psi}=1.\label{eq:DefWeakOperatorProjectionOperator}
\end{equation}
 Letting $z$ be an arbitrary complex number with complex conjugate
$z^{\conj}$, and letting $\idmatrix$ denote the identity operator
on the system's Hilbert space, define a new self-adjoint, positive-semidefinite
operator 
\begin{equation}
E\left(z\right)\defeq\left(\idmatrix+zA\right)\projector_{\Psi}\left(\idmatrix+z^{\conj}A\right)=E^{\adj}\left(z\right)\geq0,\label{eq:DefWeakOperator}
\end{equation}
 which will be called a weak operator.

Suppose now that the system's quantum state is pure and is represented
by a state vector $\ket{\Psi^{\prime}}$ that has nonvanishing inner
product with $\ket{\Psi}$, so that $\braket{\Psi^{\prime}}{\Psi}\ne0$.
Then from the general formula \eqref{eq:DvNExpectationValueFromInnerProduct}
for the expectation of an observable, the weak operator $E\left(z\right)$
has expectation value 
\begin{equation}
\expectval{E\left(z\right)}=\bra{\Psi^{\prime}}E\left(z\right)\ket{\Psi^{\prime}}=\bra{\Psi^{\prime}}\left(\idmatrix+zA\right)\ket{\Psi}\bra{\Psi}\left(\idmatrix+z^{\conj}A\right)\ket{\Psi^{\prime}}.\label{eq:WeakOperatorExpectationValue}
\end{equation}
 Simplifying, one finds 
\begin{equation}
\begin{aligned}\left.\begin{aligned}\expectval{E\left(z\right)} & =\braket{\Psi^{\prime}}{\Psi}\braket{\Psi}{\Psi^{\prime}}+z\bra{\Psi^{\prime}}A\ket{\Psi}\braket{\Psi}{\Psi^{\prime}}+z^{\conj}\bra{\Psi}A\ket{\Psi^{\prime}}\braket{\Psi^{\prime}}{\Psi}+\verts z^{2}\verts{\bra{\Psi^{\prime}}A\ket{\Psi}}^{2}\\
 & =\verts{\braket{\Psi^{\prime}}{\Psi}}^{2}\left[1+z\frac{\bra{\Psi^{\prime}}A\ket{\Psi}}{\braket{\Psi^{\prime}}{\Psi}}+z^{\conj}\frac{\bra{\Psi}A\ket{\Psi^{\prime}}}{\braket{\Psi}{\Psi^{\prime}}}+\verts z^{2}\absval{\frac{\bra{\Psi^{\prime}}A\ket{\Psi}}{\braket{\Psi^{\prime}}{\Psi}}}^{2}\right].
\end{aligned}
\right\} \end{aligned}
\label{eq:WeakOperatorExpectationValueCalculated}
\end{equation}
 In terms of the weak value $A_{w}=A_{w}\left(\Psi^{\prime},\Psi\right)$,
as defined in \eqref{eq:DefWeakValueGeneral}, one can write the expectation
value of the weak operator $E\left(z\right)$ more succinctly as
\begin{equation}
\expectval{E\left(z\right)}=\verts{\braket{\Psi^{\prime}}{\Psi}}^{2}\left[1+zA_{w}+z^{\conj}A_{w}^{\conj}+\verts z^{2}\verts{A_{w}}^{2}\right].\label{eq:WeakOperatorExpectationValueSimplified}
\end{equation}

Variously choosing the complex number $z$ to be $1$, $-1$, $i$,
and $-i$ in the formula \eqref{eq:WeakOperatorExpectationValueSimplified},
one finds 
\begin{equation}
\left.\begin{aligned}\expectval{E\left(1\right)} & =\verts{\braket{\Psi^{\prime}}{\Psi}}^{2}\left[1+2\re A_{w}+\verts{A_{w}}^{2}\right],\\
\expectval{E\left(-1\right)} & =\verts{\braket{\Psi^{\prime}}{\Psi}}^{2}\left[1-2\re A_{w}+\verts{A_{w}}^{2}\right],\\
\expectval{E\left(i\right)} & =\verts{\braket{\Psi^{\prime}}{\Psi}}^{2}\left[1-2\im A_{w}+\verts{A_{w}}^{2}\right],\\
\expectval{E\left(-i\right)} & =\verts{\braket{\Psi^{\prime}}{\Psi}}^{2}\left[1+2\im A_{w}+\verts{A_{w}}^{2}\right].
\end{aligned}
\right\} \label{eq:WeakOperatorExpectationValueOptions}
\end{equation}
 It is then a straightforward exercise to solve for $\re A_{w}$ and
$\im A_{w}$: 
\begin{equation}
\re A_{w}=\re A_{w}\left(\Psi^{\prime},\Psi\right)=\frac{\expectval{E\left(1\right)-E\left(-1\right)}}{4\verts{\braket{\Psi^{\prime}}{\Psi}}^{2}}=\frac{\expectval{\left(1/2\right)\left(A\projector_{\Psi}+\projector_{\Psi}A\right)}}{\verts{\braket{\Psi^{\prime}}{\Psi}}^{2}},\label{eq:WeakValueRePartFromExpectationValueWeakOperator}
\end{equation}
\begin{equation}
\im A_{w}=\im A_{w}\left(\Psi^{\prime},\Psi\right)=\frac{\expectval{E\left(-i\right)-E\left(i\right)}}{4\verts{\braket{\Psi^{\prime}}{\Psi}}^{2}}=\frac{\expectval{\left(1/2i\right)\left(A\projector_{\Psi}-\projector_{\Psi}A\right)}}{\verts{\braket{\Psi^{\prime}}{\Psi}}^{2}}.\label{eq:WeakValueImPartFromExpectationValueWeakOperator}
\end{equation}

These two formulas, at least in principle, provide an operational
way to extract a weak value experimentally, without any approximations,
in terms of the empirical quantities $\expectval{\left(1/2\right)\left(A\projector_{\Psi}+\projector_{\Psi}A\right)}$,
$\expectval{\left(1/2i\right)\left(A\projector_{\Psi}-\projector_{\Psi}A\right)}$,
and $\verts{\braket{\Psi^{\prime}}{\Psi}}^{2}$. That point is worth
emphasizing: weak values can be obtained, at least indirectly, through
measurements. In particular, the formulas \eqref{eq:WeakValueRePartFromExpectationValueWeakOperator}
and \eqref{eq:WeakValueImPartFromExpectationValueWeakOperator} involve
the expectation values of observables that are represented by the
curious self-adjoint combinations $\left(1/2\right)\left(A\projector_{\Psi}+\projector_{\Psi}A\right)$
and $\left(1/2i\right)\left(A\projector_{\Psi}-\projector_{\Psi}A\right)$,
which coincide with the ``flux'' and ``commutator'' operators
introduced by Cohen, Pollack (2018, equations 3 and 4)\nocite{CohenPollak:2018dowvoqouosm}.
However, neither of the two formulas \eqref{eq:WeakValueRePartFromExpectationValueWeakOperator}
and \eqref{eq:WeakValueImPartFromExpectationValueWeakOperator} \emph{as
a whole} is the expectation value of an observable, due to the presence
of the state-dependent quantity $\verts{\braket{\Psi^{\prime}}{\Psi}}^{2}=\bra{\Psi^{\prime}}\projector_{\Psi}\ket{\Psi^{\prime}}$
in each denominator.

Again, one therefore sees manifestly that a weak value, whether in
its entirety as a complex number or in terms of its real and imaginary
parts, does not correspond to a single-system observable, nor to the
kind of emergent observable that \emph{refers} to a single-system
observable, at least on the DvN axioms for quantum theory. At this
point in the discussion, the DvN axioms do not assign a weak value
\emph{any} obvious or clear physical meaning at all.

Given the preceding analysis, the null hypothesis should be that weak
values do not have any physical or metaphysical meaning, and are merely
sometimes-convenient complex-valued ratios that are amenable to extraction
via appropriate experimental protocols. To claim nonetheless that
weak values have some specific interpretation, let alone an interpretation
that makes sense at the level of a single system, would be an extraordinary
assertion, and the burden should be on the claimants to provide a
rigorous argument in favor of that view. Appeals that run afoul of
the fallacies reviewed in Subsection~\ref{subsec:Three-Relevant-Fallacies-of-Reasoning}
are insufficient for that purpose, as are appeals to the size of the
research literature on weak values.

It is worth being clear at this point on what these sorts appeals
often look like, and why they are fallacious. As explained above,
it is true that one can extract weak values indirectly through measurements
on ensembles. However, the mere fact that one can extract weak values
experimentally in this way implies nothing about whether they are
single-system observables or even whether they \emph{refer} to single-system
observables. In particular, to identify weak values as single-system
observables anyway would be to commit the ensemble fallacy \eqref{eq:DefEnsembleFallacy},
to infer anything about a quantum system or an ensemble of quantum
systems from weak values obtained via post-selection would be to commit
the post-selection fallacy \eqref{eq:DefPostSelectionFallacy}, and
to conclude anything about the physical interpretation of weak values
based solely on the fact that they can be extracted experimentally
(``If we can measure it experimentally, then how could we be wrong
about its meaning?'') would be to commit the measurementist fallacy
\eqref{eq:DefMeasurementistFallacy}.

\subsection{Weak Values for Projectors\label{subsec:Weak-Values-for-Projectors}}

Subsection~\ref{subsec:A-First-Look-at-Weak-Values} has presented
reasons for skepticism about assigning a physical significance to
weak values. One can strengthen the case for skepticism by focusing
attention on a specific class of observables: projection operators,
which arguably have the clearest meaning among all the observables
of a quantum system, at least to the extent that \emph{any} observables
of a quantum system have a clear meaning.

Consider a quantum system with an $N$-dimensional Hilbert space,
and let $\ket{c_{i}}$ for $i=1,\dots,N$ make up the orthonormal
eigenbasis for an observable $C$, so that 
\begin{equation}
C\ket{c_{i}}=c_{i}\ket{c_{i}},\qquad\braket{c_{i}}{c_{j}}=\delta_{ij}.\label{eq:DefObservableFromEigenvalueEq}
\end{equation}
 One can then define a corresponding projection-valued measure, or
PVM, by 
\begin{equation}
\projector_{i}\defeq\ket{c_{i}}\bra{c_{i}}=\projector_{i}^{\adj}=\projector_{i}^{2}.\label{eq:DefPVMFromOrthoBasis}
\end{equation}
 As a PVM, these projection operators satisfy the conditions of mutual
exclusivity \eqref{eq:DvNProjectorsMutuallyExclusive}, 
\begin{equation}
\projector_{i}\projector_{j}=\delta_{ij}\projector_{i}\label{eq:PVMMutualExclusivity}
\end{equation}
 and completeness \eqref{eq:DvNCompletenessRelation}, 
\begin{equation}
\sum_{i=1}^{N}\projector_{i}=\sum_{i=1}^{N}\ket{c_{i}}\bra{c_{i}}=\idmatrix,\label{eq:PVMCompleteness}
\end{equation}
 where $\idmatrix$ is the identity operator on the system's Hilbert
space. One can then write the observable $C$ as the spectral decomposition
\begin{equation}
C=\sum_{i=1}^{N}c_{i}\projector_{i}=\sum_{i=1}^{N}c_{i}\ket{c_{i}}\bra{c_{i}}.\label{eq:ObservableFromPVMSpectralDecomposition}
\end{equation}

Given a state vector $\ket{\Psi}$ for the system, the expectation
value \eqref{eq:DvNDefExpectationValue} of a PVM element $\projector_{i}$
has the interpretation of being the probability $p\left(c_{i}\right)$
with which a measurement of the observable $C$ will yield the specific
eigenvalue $c_{i}$: 
\begin{equation}
\expectval{\projector_{i}}=\bra{\Psi}\projector_{i}\ket{\Psi}=p\left(c_{i}\right).\label{eq:ExpectationValuePVMAsProbability}
\end{equation}
 Accordingly, this expectation value $\expectval{\projector_{i}}$
is always real-valued, always lies between the extreme values $0$
and $1$, and satisfies the usual normalization constraint,  in virtue
of the completeness relation \eqref{eq:PVMCompleteness}: 
\begin{equation}
\sum_{i=1}^{N}p\left(c_{i}\right)=\sum_{i=1}^{N}\expectval{\projector_{i}}=\bra{\Psi}\sum_{i=1}^{N}\projector_{i}\ket{\Psi}=\bra{\Psi}\idmatrix\ket{\Psi}=1.\label{eq:NormalizationConstraintFromPVMCompleteness}
\end{equation}
 One then interprets the equation $\expectval{\projector_{i}}=1$
as assigning the value \textquoteleft true\textquoteright{} to the
proposition ``A measurement of $C$ will yield the eigenvalue $c_{i}$,''
and the equation $\expectval{\projector_{i}}=0$ as assigning the
value \textquoteleft false\textquoteright{} to that same proposition.

These interpretative assertions are consistent with the expectation
value $\expectval C$ of $C$, given the spectral decomposition \eqref{eq:ObservableFromPVMSpectralDecomposition}.
Indeed, 
\begin{equation}
\bra{\Psi}C\ket{\Psi}=\sum_{i=1}^{N}c_{i}\bra{\Psi}\projector_{i}\ket{\Psi}=\sum_{i=1}^{N}c_{i}p\left(c_{i}\right)=\expectval C,\label{eq:ExpectationValueObservableFromPVM}
\end{equation}
 in line with the definition \eqref{eq:DvNDefExpectationValue} of
an expectation value.

These interpretative assertions and relationships break down for weak
values. It is certainly true that the weak values $\projector_{i,w}$
sum to $1$, a feature that bears a resemblance to the normalization
constraint \eqref{eq:NormalizationConstraintFromPVMCompleteness}:
\begin{equation}
\sum_{i=1}^{N}\projector_{i,w}=\frac{\bra{\Psi^{\prime}}\sum_{i}\projector_{i}\ket{\Psi}}{\braket{\Psi^{\prime}}{\Psi}}=\frac{\bra{\Psi^{\prime}}\idmatrix\ket{\Psi}}{\braket{\Psi^{\prime}}{\Psi}}=1.\label{eq:SumPVMWeakValuesEq1}
\end{equation}
 However, each individual weak value $\projector_{i,w}$ can be essentially
any complex number, with unbounded modulus, so that $0$ and $1$
are no longer extreme values. As acknowledged by Vaidman (1996, p.
903, equation 10)\nocite{Vaidman:1996wmeor}, the weak value of a
projection operator can even be $-1$. Any connection between projection
operators and probabilities (even conditional probabilities) is then
utterly lost, as is any reason to say that $\projector_{i,w}=1$
should refer to anything being \textquoteleft true\textquoteright{}
or that $\projector_{i,w}=0$ should refer to anything being \textquoteleft false.\textquoteright{}

As the present work will review, especially in Subsection~\ref{subsec:Historical-Approaches-to-Interpretation-of-Weak-Values},
a common view in the research literature is to assert that a weak
value $A_{w}=A_{w}\left(\Psi^{\prime},\Psi\right)$ should somehow
be regarded as the value that the observable $A$ has for an ensemble
of quantum systems pre-selected to have the initial state vector $\ket{\Psi}$
and post-selected to have a post-measurement state vector $\ket{\Psi^{\prime}}$.
(See, for example, Aharonov, Vaidman 1990.)\nocite{AharonovVaidman:1990poaqsdttibtm}
What could it possibly mean for a quantum system to have a value for
a projection operator that is given by a complex number lying far
outside the unit circle from the origin of the complex plane?

Looking back at the weak operator defined in \eqref{eq:DefWeakOperator}
offers little help. For $A=\projector_{i}$, one has weak operators
defined by 
\begin{equation}
E_{i}\left(z\right)\defeq\left(\idmatrix+z\projector_{i}\right)\projector_{\Psi}\left(\idmatrix+z^{\conj}\projector_{i}\right).\label{eq:DefWeakOperatorsForPVM}
\end{equation}
 The self-adjoint combinations appearing in \eqref{eq:WeakValueRePartFromExpectationValueWeakOperator}
and \eqref{eq:WeakValueImPartFromExpectationValueWeakOperator} then
become $\left(1/2\right)\left(\projector_{i}\projector_{\Psi}+\projector_{\Psi}\projector_{i}\right)$
and $\left(1/2i\right)\left(\projector_{i}\projector_{\Psi}-\projector_{\Psi}\projector_{i}\right)$.
At least on the DvN axioms, there is no sense in which the first combination
having the expectation value $1\times\verts{\braket{\Psi^{\prime}}{\Psi}}^{2}$
and the second having the expectation value $0$, so that $\projector_{i,w}=\re\projector_{i,w}+i\im\projector_{i,w}=1$,
should imply anything about whether any property of the quantum system
is \textquoteleft true.\textquoteright{} Nor is there any sense in
which both combinations having expectation value $0$, so that $\projector_{i,w}=\re\projector_{i,w}+i\im\projector_{i,w}=0$,
should imply that anything about the quantum system is \textquoteleft false.\textquoteright{}

These observations directly challenge statements made, for instance,
by Aharonov, Popescu, Rohrlich, Skrzypczyk (2013, pp. 5\textendash 6,
equations 6\textemdash 9)\nocite{AharonovPopescuRohrlichSkrzypczyk:2013qcc}.
The present work will address some of the downstream implications
of these arguments in Subsection~\ref{subsec:Quantum-Cheshire-Cats}.

\subsection{The AAV Experimental Protocol\label{subsec:The-AAV-Experimental-Protocol}}

The AAV paper (Aharonov, Albert, Vaidman 1988)\nocite{AharonovAlbertVaidman:1988htroamoacotsoas12pctotb1}
presented a practical, if approximate, means of extracting weak values
by a somewhat complicated procedure that involves a subtle interplay
of entanglement, weak measurements, and ensembles obtained through
post-selection. The AAV experimental protocol will be presented here,
both to connect the present work with the AAV paper more comprehensively,
and also to make clear why the AAV experimental protocol does not,
in the end, lead to a more perspicuous interpretation of weak values,
despite what one might have hoped.

 The derivation to follow will be similar to the derivation of the
ABL rule presented in other work  (Barandes 2026)\nocite{Barandes:2026taratpops},
but will be fully self-contained. 

To begin, one considers three quantum systems: a subject system to
be studied, a measuring device, and an external agent or observer.
The total system's Hilbert space is given by the tensor product 
\begin{equation}
\hilbspace_{\textrm{tot}}=\hilbspace\tensorprod\hilbspace_{\textrm{dev}}\tensorprod\hilbspace_{\textrm{obs}},\label{eq:WeakValuesTotalHilbertSpaceSubjectDeviceObs}
\end{equation}
 where $\hilbspace$, $\hilbspace_{\textrm{dev}}$, and $\hilbspace_{\textrm{obs}}$
are the respective Hilbert spaces for the subject system, the measuring
device, and the observer.

As with the ABL rule, one supposes that for the initial state vector
$\ket{\Psi_{\textrm{tot}}}$ of the total system, the subject system
has a generic initial state vector $\ket{\Psi}$, and the observer
is assigned its usual \textquoteleft ready\textquoteright{} state
vector $\ket{\textrm{obs}\left(\emptyset\right)}$. The measuring
device, by contrast, is assumed to begin with a specially chosen state
vector $\ket{\Phi_{\textrm{dev}}}$, to be defined momentarily. Altogether,
the initial state vector for the total system is 
\begin{equation}
\ket{\Psi_{\textrm{tot}}}=\ket{\Psi}\tensorprod\ket{\Phi_{\textrm{dev}}}\tensorprod\ket{\textrm{obs}\left(\emptyset\right)}.\label{eq:WeakValuesTotalStateVectorInitial}
\end{equation}
 The overall plan will be to pick an observable $A$ belonging to
the subject system, together with a second state vector $\ket{\Psi^{\prime}}$
for the subject system to be used for post-selection, and then indirectly
obtain the weak value $A_{w}\left(\Psi^{\prime},\Psi\right)$ corresponding
to $A$, $\ket{\Psi}$, and $\ket{\Psi^{\prime}}$, as defined in
\eqref{eq:DefWeakValueGeneral}

One assumes here that the measuring device has a canonical pair of
observables, $q$ and $p$, where the momentum observable $p$ is
regarded as the measuring device's pointer variable, and $q$ is its
conjugate coordinate.\footnote{Note that Bohm (1952)\nocite{Bohm:1952siqtthvii}, Vaidman (1996a)\nocite{Vaidman:1996wmeor},
and Dressel et al. (2014)\nocite{DresselMalikMiattoJordanBoyd:2014cuqwvbaa}
choose a convention in which the measuring device's canonical coordinate
$q$, rather than its canonical momentum $p$, is taken to be the
pointer variable. For that convention, one should replace $q$ with
$p$ in the interaction Hamiltonian \eqref{eq:DefWeakValuesHamiltonian},
and the overall minus sign should be removed.} For a standard von Neumann measurement (von Neumann 1932)\nocite{vonNeumann:1932mgdq},
the measurement outcome would ordinarily be read directly off of the
pointer variable $p$, after imposing a collapse \eqref{eq:DvNCollapse}
to the total system's state vector. That procedure is modified for
the experimental extraction of weak values.

From the assumption that $p$ is the measuring device's pointer variable,
one assumes that it has the same measurement units as $A$, in which
case the product $qA$ has units of the (reduced) Planck constant
$\hbar$.

One takes the initial state vector $\ket{\Phi_{\textrm{dev}}}$ for
the measuring device to have a narrow Gaussian profile in coordinate
space, centered at $q=0$. That is, the measuring device's coordinate-space
wave function is assumed to have the form 
\begin{equation}
\braket q{\Phi_{\textrm{dev}}}\propto e^{-q^{2}/4\sigma_{q}^{2}},\label{eq:WeakValuesInitialCoordSpaceWaveFunctionDevice}
\end{equation}
 so that the corresponding coordinate-space probability distribution
is 
\begin{equation}
\verts{\braket q{\Phi_{\textrm{dev}}}}^{2}\propto e^{-q^{2}/2\sigma_{q}^{2}},\label{eq:WeakValuesInitialCoordSpaceProbabilityDistributionDevice}
\end{equation}
 where the standard deviation $\sigma_{q}$ is assumed to be very
small relative to the reciprocal spacing between the eigenvalues of
the observable $A$, up to a factor of $\hbar$. This assumed initial
state vector then has a very broad Gaussian profile in momentum space,
\begin{equation}
\braket p{\Phi_{\textrm{dev}}}\propto e^{-p^{2}/4\sigma_{p}^{2}},\label{eq:WeakValuesInitialMomentumSpaceWaveFunctionDevice}
\end{equation}
 with a standard deviation $\sigma_{p}=\hbar/2\sigma_{q}$ that is
very large compared with the spacing between the eigenvalues of $A$.

This broadness of the measuring device's initial quantum state in
momentum space is understood to mean that the measuring device will
effectively interact only very weakly with the subject system, so
the interaction is accordingly called a \textquoteleft weak measurement\textquoteright{}
(Aharonov, Albert, Casher, Vaidman 1987)\nocite{AharonovAlbertCasherVaidman:1987sqe}.
The analysis ahead will show explicitly that a weak measurement perturbs
the quantum state of the measuring device only in a slight, predictable
way, and leaves the quantum state of the subject system essentially
unchanged, at least for most practical purposes.

Ignoring any other time evolution in the system, under the assumption
that the interaction between the subject system and the measuring
device occurs over a very short time interval, and following von
Neumann (1932)\nocite{vonNeumann:1932mgdq}, one takes the interaction
Hamiltonian to be\footnote{In the original German: ``...dann bleibt von $H$ nur der für die
Messung entscheidende Wechselwirkungs-Energieanteil übrig. Für diesen
wählen wir die besondere Form $\frac{h}{2\pi i}q\frac{\partial}{\partial r}$''
(von Neumann 1932, Section VI.3, p. 236)\nocite{vonNeumann:1932mgdq}.} 

\begin{equation}
H\left(t\right)=-g\left(t\right)qA.\label{eq:DefWeakValuesHamiltonian}
\end{equation}
 Here $g\left(t\right)$ is a function that plays the role of a coupling
parameter, has measurement units of inverse-time, has compact support
that is sharply peaked at the interaction time, and is normalized
to unity: 
\begin{equation}
\int dt\,g\left(t\right)=1.\label{eq:WeakValuesNormalizedPulseFunction}
\end{equation}

Choosing an arbitrary orthonormal basis for the subject system labeled
by $b$, the total system then has the following time evolution:
\begin{equation}
\left.\begin{aligned}\ket{\Psi_{\textrm{tot}}} & =\ket{\Psi}\tensorprod\ket{\Phi_{\textrm{dev}}}\tensorprod\ket{\textrm{obs}\left(\emptyset\right)}\\
 & \mapsto e^{-\left(i/\hbar\right)\int dt\,H\left(t\right)}\ket{\Psi}\tensorprod\ket{\Phi_{\textrm{dev}}}\tensorprod\ket{\textrm{obs}\left(\emptyset\right)}\\
 & =\sum_{b}\int dq\,\left[\ket b\tensorprod\ket q\right]\left[\bra b\tensorprod\bra q\right]e^{-\left(i/\hbar\right)\int dt\,H\left(t\right)}\left[\ket{\Psi}\tensorprod\ket{\Phi_{\textrm{dev}}}\right]\tensorprod\ket{\textrm{obs}\left(\emptyset\right)}.
\end{aligned}
\right\} \label{eq:WeakValuesFirstStepInteraction}
\end{equation}
 Thus, one finds an inner product that needs to be calculated, involving
just the subject system and the measuring device: 
\begin{equation}
\left[\bra b\tensorprod\bra q\right]e^{-\left(i/\hbar\right)\int dt\,H\left(t\right)}\left[\ket{\Psi}\tensorprod\ket{\Phi_{\textrm{dev}}}\right]=\bra be^{iqA/\hbar}\ket{\Psi}\braket q{\Phi_{\textrm{dev}}}.\label{eq:WeakValuesInnerProduct}
\end{equation}

Because the measuring device is assumed to have a coordinate-space
profile $\braket q{\Phi_{\textrm{dev}}}\propto\exp\left(-q^{2}/4\sigma_{q}^{2}\right)$
in \eqref{eq:WeakValuesInnerProduct} that is sharply peaked at $q=0$,
one can effectively treat $q$ as a small parameter and therefore
approximate the exponential operator as the first-order term in its
Taylor series: 
\begin{equation}
e^{iqA/\hbar}\approx1+\frac{iqA}{\hbar}.\label{eq:WeakValueApproxExponentiatedOperator}
\end{equation}
 Hence, 
\begin{equation}
\left.\begin{aligned}\bra be^{iqA/\hbar}\ket{\Psi} & \approx\bra b\left(1+\frac{iqA}{\hbar}\right)\ket{\Psi}\\
 & =\braket b{\Psi}+\frac{iq}{\hbar}\bra bA\ket{\Psi}\\
 & =\braket b{\Psi}\left(1+\frac{iqA_{w}}{\hbar}\right)\\
 & \approx\braket b{\Psi}e^{iqA_{w}/\hbar},
\end{aligned}
\right\} \label{eq:WeakValuesCalculationInnerProductPrefactor}
\end{equation}
  where the weak value $A_{w}$ is defined as in \eqref{eq:DefWeakValueGeneral}
by 
\begin{equation}
A_{w}=A_{w}\left(b,\Psi\right)\defeq\frac{\bra bA\ket{\Psi}}{\braket b{\Psi}}\qquad\left[\textrm{compare to AAV's eq. }\left(6\right)\right],\label{eq:DefWeakValueAAV}
\end{equation}
 and, again, is generically complex-valued.  Hence, the inner product
\eqref{eq:WeakValuesInnerProduct} reduces approximately to 
\begin{equation}
\braket b{\Psi}e^{iqA_{w}/\hbar}\braket q{\Phi_{\textrm{dev}}}\approx\braket b{\Psi}\braket q{\Phi_{\textrm{dev}\given A_{w}}},\label{eq:WeakValuesInnerProductResult}
\end{equation}
 where the updated state vector $\ket{\Phi_{\textrm{dev}\given A_{w}}}$
of the measuring device has coordinate-space wave function 
\begin{equation}
\left.\begin{aligned}\braket q{\Phi_{\textrm{dev}\given A_{w}}} & =e^{iqA_{w}/\hbar}\braket q{\Phi_{\textrm{dev}}}\\
 & \propto e^{-q^{2}/4\sigma_{q}^{2}+iqA_{w}/\hbar}.
\end{aligned}
\right\} \label{eq:WeakValuesFinalCoordSpaceWaveFunctionDevice}
\end{equation}

Thus, the total system's state vector \eqref{eq:WeakValuesFirstStepInteraction}
becomes approximately 
\begin{equation}
\sum_{b}\int dq\,\braket b{\Psi}\braket q{\Phi_{\textrm{dev}\given A_{w}\left(b,\Psi\right)}}\ket b\tensorprod\ket q\tensorprod\ket{\textrm{obs}\left(\emptyset\right)},\label{eq:WeakValuesTotalStateVectorAfterInteraction}
\end{equation}
 where the explicit notation $A_{w}\left(b,\Psi\right)$ is intended
to highlight the entanglement between the subject system and the measuring
device.

It is important to note that the orthonormal basis labeled by $b$
here is totally arbitrary at this point. No physical feature of the
subsystems or the interaction between the measuring device and the
subject system has singled out $b$ or depends on $b$. The measuring
device may have been perturbed by the interaction, but not yet in
a way that actually depends on the specific weak values $A_{w}\left(b,\Psi\right)$,
for the specific basis labeled by $b$.

The observer then carries out a standard projective measurement on
the \emph{subject system} in the basis labeled by $b$, thereby yielding
the following final state vector for the total system, which now features
entanglement between all three subsystems: 
\begin{equation}
\ket{\Psi_{\textrm{tot}}^{\prime}}=\sum_{b}\int dq\,\braket b{\Psi}\braket q{\Phi_{\textrm{dev}\given A_{w}\left(b,\Psi\right)}}\ket b\tensorprod\ket q\tensorprod\ket{\textrm{obs}\left(b\right)}.\label{eq:WeakValuesTotalStateVectorAfterObsMeasurement}
\end{equation}
 Here $\textrm{obs\ensuremath{\left(b\right)}}$ indicates that the
observer has obtained the measurement outcome $b$.

\emph{Crucially}, it is only at the conclusion of \emph{this step}
that the measuring device's quantum state definitively gains a dependence
on the weak values $A_{w}\left(b,\Psi\right)$, for the specific orthonormal
basis labeled by $b$. That dependence on $A_{w}\left(b,\Psi\right)$
did not arise merely from the interaction between the measuring device
and the subject system, but required the projective measurement by
the observer in the basis labeled by $b$.

Next, from among the possible final measurement results labeled by
$b$, the observer post-selects a \emph{specific} choice $b=\Psi^{\prime}$.
This measurement and post-selection induces a DvN collapse of the
total system's state vector, thereby breaking the entanglement and
yielding\footnote{As Aharonov and Vaidman note in a separate paper (Aharonov, Vaidman
1990)\nocite{AharonovVaidman:1990poaqsdttibtm}, the observer could
instead measure the measuring device's pointer variable $p$ or its
conjugate coordinate $q$ first, and then, secondarily, measure the
subject system's label $b$ and post-select on $b=\Psi^{\prime}$,
without any meaningful difference to the final state vector or statistics.
Whatever the order of these two final steps, it is only due to the
observer's choice of $b$ and post-selection on $b=\Psi^{\prime}$
that any dependence arises on a specific weak value $A_{w}\left(b,\Psi\right)=A_{w}\left(\Psi^{\prime},\Psi\right)$.} 
\begin{equation}
\ket{\Psi_{\textrm{tot}}^{\prime}}=\ket{\Psi^{\prime}}\tensorprod\left[\int dq\,\braket q{\Phi_{\textrm{dev}\given A_{w}\left(\Psi^{\prime},\Psi\right)}}\ket q\right]\tensorprod\ket{\textrm{obs}\left(\Psi^{\prime}\right)}.\label{eq:WeakValuesTotalStateVectorAfterObsMeasurementAfterPostselection}
\end{equation}

Due to the observer's projective measurement and post-selection, the
updated quantum state for the measuring device alone now definitively
depends on the specific weak value $A_{w}=A_{w}\left(\Psi^{\prime},\Psi\right)$,
and is given by 
\begin{equation}
\left.\begin{aligned}\ket{\Phi_{\textrm{dev}\given A_{w}}} & =\int dq\,\ket q\braket q{\Phi_{\textrm{dev}\given A_{w}}}\\
 & \propto\int dq\,\ket qe^{-q^{2}/4\sigma_{q}^{2}+iqA_{w}/\hbar}\\
 & \propto\int dp\,\ket pe^{-\left(p-A_{w}\right)^{2}/4\sigma_{p}^{2}}.
\end{aligned}
\right\} \label{eq:WeakValuesFinalStateVectorDevice}
\end{equation}
 Decomposing the weak value $A_{w}=\re A_{w}+i\im A_{w}$ into its
real and imaginary parts, and carrying out a straightforward calculation,
one can show that the probability distribution in momentum space turns
out to be 
\begin{equation}
\verts{\braket p{\Phi_{\textrm{dev}\given A_{w}}}}^{2}\propto e^{-\left(p-\re A_{w}\right)^{2}/2\sigma_{p}^{2}},\label{eq:WeakValuesDevProbabilityDistributionMomentumSpace}
\end{equation}
 whereas the probability distribution in coordinate space is 
\begin{equation}
\verts{\braket q{\Phi_{\textrm{dev}\given A_{w}}}}^{2}\propto e^{-\left(q+2\sigma_{q}^{2}\im A_{w}/\hbar\right)^{2}/2\sigma_{q}^{2}}.\label{eq:WeakValuesDevProbabilityDistributionCoordSpace}
\end{equation}

By assumption, the standard deviation $\sigma_{p}$ of the momentum-space
probability distribution \eqref{eq:WeakValuesDevProbabilityDistributionMomentumSpace}
for the measuring device is very large, so the peak around the expectation
value $\expectval p=\re A_{w}$ is extremely broad, thereby necessitating
a large ensemble to gain experimental access to $\re A_{w}$. Although
the standard deviation $\sigma_{q}=\hbar/2\sigma_{p}$ of the coordinate-space
probability distribution \eqref{eq:WeakValuesDevProbabilityDistributionCoordSpace}
is, by contrast, very small, meaning that the peak around the expectation
value $\expectval q=-2\sigma_{q}^{2}\im A_{w}/\hbar=-\hbar\,\im A_{w}/2\sigma_{p}^{2}$
is very narrow, this expectation value itself is extremely small,
and so, again, would require a large ensemble to determine experimentally.

These conclusions hold even if, say, the observable $A$ in question
is a projector $\projector=\projector^{\adj}=\projector^{2}$ representing
a true-or-false question, and even if, furthermore, its weak value
happens to be $\projector_{w}=1$ or $\projector_{w}=0$. As explained
in Subsection~\ref{subsec:Weak-Values-for-Projectors}, there is,
in particular, no sense in which the ratio $\projector_{w}=\bra{\Psi^{\prime}}\projector\ket{\Psi}/\braket{\Psi^{\prime}}{\Psi}$
having the value $1$ means that anything about the subject system
is definitively \textquoteleft true,\textquoteright{} or having the
value $0$ means that anything about the subject system is definitively
\textquoteleft false,\textquoteright{} again in contrast with claims
made by Aharonov, Popescu, Rohrlich, Skrzypczyk (2013, pp. 5\textendash 6,
equations 6\textendash 9)\nocite{AharonovPopescuRohrlichSkrzypczyk:2013qcc},
as noted in Subsection~\ref{subsec:Weak-Values-for-Projectors}.

In principle, then, to obtain a weak value like $A_{w}$, one sets
up a large physical ensemble of $N\gg1$ identically prepared bipartite
systems. Each bipartite system consists of a subject system with an
observable $A$ and initially represented by a state vector $\ket{\Psi}$,
together with a measuring device prepared in a narrow Gaussian in
the space of the coordinate conjugate to its pointer variable. One
then lets the two subsystems in each bipartite system suitably interact.
At the end of the whole experiment, an external agent or observer
carries out a projective measurement on each subject system and keeps
only the bipartite systems for which the subject system is found to
have some specified final state vector $\ket{\Psi^{\prime}}$, thereby
deliberately implementing a \textquoteleft post-selection\textquoteright{}
that projects the subject system's quantum state down to $\ket{\Psi^{\prime}}$.
The observer then carries out a projective measurement on each measuring
device itself\textemdash specifically, either a measurement of the
measuring device's pointer variable $p$ or its conjugate coordinate
$q$. Having carried out this overall experimental protocol on a large
physical ensemble $N\gg1$ of these bipartite systems, and after computing
the relevant statistical means, the external agent can obtain an estimate
of the real and imaginary parts of the weak value $A_{w}$.

Despite predating the AAV paper, the AACV paper considered the slightly
more general possibility of implementing the extraction of more than
one weak value in the course of the overall experimental protocol.
If each measuring device has multiple pointer variables $p_{1},\dots,p_{n}$,
with corresponding conjugate coordinates $q_{1},\dots,q_{n}$, then
one can extract weak values in succession on the same subject system,
all from the same pre-selected state vector $\ket{\Psi}$ and the
same post-selected state vector $\ket{\Psi^{\prime}}$. For this purpose,
one replaces the original Hamiltonian \eqref{eq:DefWeakValuesHamiltonian}
with 
\begin{equation}
\sum_{\alpha=1}^{n}H_{\alpha}\left(t\right)=-\sum_{\alpha=1}^{n}g_{\alpha}\left(t\right)q_{\alpha}A_{\alpha},\label{eq:DefWeakValuesHamiltonianMultiObservable}
\end{equation}
 and one assumes that the measuring device's coordinate-space profile
is a Gaussian that is sharply peaked at $\left(q_{1},\dots,q_{n}\right)=\left(0,\dots,0\right)$:
\begin{equation}
\braket{q_{1},\dots,q_{n}}{\Phi_{\textrm{dev}}}\propto e^{-\sum_{\alpha}q_{\alpha}^{2}/4\sigma_{q_{\alpha}}^{2}}.\label{eq:WeakValuesInitialCoordSpaceWaveFunctionDeviceMultiObservable}
\end{equation}
In place of the inner product \eqref{eq:WeakValuesInnerProduct},
one now has 
\begin{equation}
\left[\bra b\tensorprod\bra{q_{1},\dots,q_{n}}\right]e^{-\left(i/\hbar\right)\sum_{\alpha}\int dt\,H_{\alpha}\left(t\right)}\left[\ket{\Psi}\tensorprod\ket{\Phi_{\textrm{dev}}}\right]=\bra be^{i\sum_{\alpha}q_{\alpha}A_{\alpha}/\hbar}\ket{\Psi}\braket{q_{1},\dots,q_{n}}{\Phi_{\textrm{dev}}}.\label{eq:WeakValuesInnerProductMultipleMeasurements}
\end{equation}
 Treating the coordinates $q_{1},\dots,q_{n}$ as effectively small
parameters\textemdash a crucial step for treating the interactions
as weak measurements\textemdash one then obtains 
\begin{equation}
\bra be^{i\sum_{\alpha}q_{\alpha}A_{\alpha}/\hbar}\ket{\Psi}\approx\bra b\left(1+\frac{i}{\hbar}\sum_{\alpha=1}^{n}q_{\alpha}A_{\alpha}\right)\ket{\Psi}\approx\braket b{\Psi}e^{i\sum_{\alpha}q_{\alpha}A_{\alpha,w}/\hbar},\label{eq:WeakValuesCalculationInnerProductPrefactorMultiObservable}
\end{equation}
 so 
\begin{equation}
\bra be^{i\sum_{\alpha}q_{\alpha}A_{\alpha}/\hbar}\ket{\Psi}\braket{q_{1},\dots,q_{n}}{\Phi_{\textrm{dev}}}\approx\braket b{\Psi}\braket{q_{1},\dots,q_{n}}{\Phi_{\textrm{dev}\given A_{1,w},\dots,A_{n,w}}}.\label{eq:WeakValuesInnerProductResultMultiObservable}
\end{equation}
 Here the measuring device's updated coordinate-space wave function
generalizes \eqref{eq:WeakValuesFinalCoordSpaceWaveFunctionDevice}
to 
\begin{equation}
\braket{q_{1},\dots,q_{n}}{\Phi_{\textrm{dev}\given A_{1,w},\dots,A_{n,w}}}\propto e^{-\sum_{\alpha}\left(q_{\alpha}^{2}/4\sigma_{q_{\alpha}}^{2}+iq_{\alpha}A_{\alpha,w}/\hbar\right)}.\label{eq:WeakValuesFinalCoordSpaceWaveFunctionDeviceMultipleMeasurements}
\end{equation}
 After post-selection to $b=\Psi^{\prime}$, the weak values are,
in keeping with \eqref{eq:DefWeakValueGeneral}, given by 
\begin{equation}
A_{\alpha,w}\equiv\frac{\bra{\Psi^{\prime}}A_{\alpha}\ket{\Psi}}{\braket{\Psi^{\prime}}{\Psi}}=A_{\alpha,w}\left(\Psi^{\prime},\Psi\right)\quad\left[\alpha=1,\dots,n\right].\label{eq:DefWeakValuesMultipleMeasurements}
\end{equation}
 These weak values are all independent of each other, and so are not
subject to conditions like the uncertainty principle. They also satisfy
a form of robustness or replicability, in the sense that if the same
observable is subjected to a weak-value measurement twice, then the
corresponding weak values will be the same: 
\begin{equation}
A_{\alpha}=A_{\beta}\implies A_{\alpha,w}=A_{\beta,w}.\label{eq:WeakValueRobustness}
\end{equation}

\subsection{The Interpretation of Weak Values\label{subsec:The-Interpretation-of-Weak-Values}}

A weak value $A_{w}$, as defined in \eqref{eq:DefWeakValueGeneral},
manifestly involves a renormalized matrix element $\bra{\Psi^{\prime}}A\ket{\Psi}/\braket{\Psi^{\prime}}{\Psi}$
of the self-adjoint operator $A$. Being able to calculate matrix
elements of self-adjoint operators through an ensemble-based experimental
protocol is clearly a useful technique on its own terms, again as
reviewed by Dressel et al. (2014)\nocite{DresselMalikMiattoJordanBoyd:2014cuqwvbaa},
but one might ask for more. In particular, given the discussion in
Subsection~\ref{subsec:A-First-Look-at-Weak-Values} and the review
of the AAV experimental protocol in Subsection~\ref{subsec:The-AAV-Experimental-Protocol},
one might reasonably wonder if weak values have a more profound physical
or metaphysical significance.

However, as explained in Subsection~\ref{subsec:A-First-Look-at-Weak-Values},
the DvN axioms identify a weak value only as an irreducibly emergent
observable of an \emph{ensemble} of systems, and not even the kind
of ensemble observable that \emph{refers} to a single-system observable.
Nothing about the AAV experimental protocol changes that basic fact.

It is true that in the derivation presented in Subsection~\ref{subsec:The-AAV-Experimental-Protocol},
the final state vector $\ket{\Phi_{\textrm{dev}\given A_{w}}}$ of
each measuring device, as expressed in \eqref{eq:WeakValuesFinalStateVectorDevice},
contains a dependence on the weak value $A_{w}=A_{w}\left(\Psi^{\prime},\Psi\right)$.
However, merely recording an ensemble observable's value into the
internal memory of a single measuring device does not make that ensemble
observable into a single-system observable, or else \emph{all} ensemble
observables would be single-system observables. What matters is not
the system into which the observable's value is \emph{stored}, but
the system from which the observable is \emph{obtained}. Generic weak
values, just like DvN measurement probabilities and expectation values,
are \emph{obtained} from ensembles of subject systems, and not from
any one subject system.

It might be tempting to argue that weak values must have some sort
of single-system meaning based on the fact that the AAV experimental
protocol involves measurement-like interactions between the subject
systems and the measuring devices, together with the fact that these
measurement-like interactions perturb the quantum states of the measuring
devices (Vaidman 1996a)\nocite{Vaidman:1996wmeor}. If a procedure
perturbs the quantum states of measuring devices, then how can whatever
produces the perturbation fail to exist in some physical sense? The
question, however, is about whether or not whatever produces the perturbation
is a single-system feature or property, and, moreover, whether or
not it is a statistical artifact of the choice of post-selection.

One should keep in mind that the quantum states of the measuring devices
did \emph{not} develop their dependence on the weak value $A_{w}=A_{w}\left(\Psi^{\prime},\Psi\right)$
due only to the interactions between the measuring devices and the
subject systems. As emphasized explicitly in Subsection~\ref{subsec:The-AAV-Experimental-Protocol}
immediately before and after \eqref{eq:WeakValuesTotalStateVectorAfterObsMeasurement},
that dependence on $A_{w}=A_{w}\left(\Psi^{\prime},\Psi\right)$ was
the result of the external observer carrying out a final projective
measurement and post-selection on the subject systems for the specific
choice of state vector $\ket{\Psi^{\prime}}$. Those decisions and
actions taken by the observer were what imbued the quantum states
of the measuring devices with a dependence on $\ket{\Psi^{\prime}}$,
and, consequently, with a dependence on the weak value $A_{w}=A_{w}\left(\Psi^{\prime},\Psi\right)$.

To argue that the weak value $A_{w}$ was just there, waiting to be
found, would be akin to arguing that a sculpture already exists in
a block of marble and merely needs to be revealed, as imagined by
Robert Allyn in this paper's epigraph.\footnote{Variants of this quotation are sometimes attributed to Michelangelo.}

In particular, attempting to identify a weak value obtained from the
AAV experimental protocol as directly implying some value for a single-system
observable, or neglecting the manifest dependence of a weak value
on the arbitrary choice of post-selection by the observer, would be
to engage in the sort of fallacious reasoning discussed in Subsection~\ref{subsec:Three-Relevant-Fallacies-of-Reasoning}.

An example might be helpful here. Forgetting about quantum theory
for a moment, consider a large ensemble of classical pendulums, each
one adjacent to its own individual detector and electronic memory-storage
device. Suppose that each detector measures the amplitude, frequency,
and phase of its nearby pendulum, so that the detector's memory-storage
device develops a classical correlation with that nearby pendulum.
Suppose also that an external experimenter has the ability to observe
the pendulums directly, bypassing the detectors.

By a judicious choice of post-selection directly on the pendulums,
the experimenter can cull the overall ensemble down to a subensemble
of pendulums whose amplitudes, frequencies, and phases compose the
Fourier series for Bach's Toccata and Fugue. As a side-effect, the
memory-storage devices in this subensemble\textemdash which, again,
are each correlated with their nearby pendulum\textemdash would then
collectively contain the Toccata and Fugue, as a Fourier series. This
experimental protocol would therefore provide an indirect way to encode
the Toccata and Fugue into memory-storage devices for various practical
uses.

It would obviously be incorrect, however, to take this result to imply
anything about the individual properties of the pendulums, let alone
to ascribe a small amount of \textquoteleft Toccata-and-Fugue-ness\textquoteright{}
(\textquoteleft fugacity\textquoteright ?) to each pendulum. Bach's
Toccata and Fugue here is clearly an emergent ensemble observable,
and a statistical artifact of the experimenter's choice of post-selection
on the pendulums, rather than a reflection of a property intrinsic
to each individual pendulum itself. To confuse this ensemble observable
with a single-system property of each pendulum, or to use this ensemble
observable to make direct inferences about any single-system property
of each pendulum, would entail committing the ensemble fallacy \eqref{eq:DefEnsembleFallacy}.
Moreover, to draw an erroneous conclusion about the pendulums based
on the use of post-selection would mean committing the post-selection
fallacy \eqref{eq:DefPostSelectionFallacy}. Finally, to attempt to
evade these two charges of fallacious reasoning merely by pointing
out that the Toccata and Fugue can, in fact, be extracted experimentally
from this protocol (``How could the result be mistaken if we can
get it from an experiment?'') would be to commit the measurementist
fallacy \eqref{eq:DefMeasurementistFallacy}.

Granted, the foregoing example is, by assumption, classical. However,
if one wanted to construct a defense of conventional interpretations
of weak values against the charges of fallacious reasoning presented
in this paper, it would not be enough merely to point out that weak
values arise from quantum-mechanical experiments, and to say that
quantum systems are different from classical systems. Of course it
is true that quantum systems are different from classical systems.
Quantum systems are, after all, characterized by the DvN axioms (or
axioms that at least make empirically similar claims), with exotic-looking
downstream implications that appear to be conceptually distinct from
the sorts of phenomena that one finds for classical systems. However,
a proper defense of conventional interpretations of weak values would
require carefully showing that quantum theory differs from classical
physics in the \emph{right kinds of ways} that make \emph{the right
kind of difference}. Indeed, putting aside the classical example presented
above, the criticisms laid out in the present work and the fallacies
listed in Subsection~\eqref{subsec:Three-Relevant-Fallacies-of-Reasoning}
are not based in classical physics, but take weak values seriously
from a quantum-mechanical perspective.

\subsection{Historical Approaches to the Interpretation of Weak Values\label{subsec:Historical-Approaches-to-Interpretation-of-Weak-Values}}

In a 2007 paper, Wiseman referred to a weak value as a kind of \textquoteleft mean\textquoteright{}
value, purportedly akin to the expectation value of a single-system
observable: 
\begin{quotation}
Consider a weak measurement of some observable $a$. A \emph{weak
value}, denoted $\langle\hat{a}_{w}\rangle_{\ket{\psi}}$, is the
\emph{mean} value from such weak measurements on an ensemble of systems,
each prepared in the state $\ket{\psi}$. So far, the weak value is
no different from the strong value, the ensemble mean value of strong
or precise measurements. However, the weak value differs from the
strong value if one calculates the mean from a subensemble obtained
by post-selecting only those results for which a later strong measurement
reveals the system to be in state $\ket{\phi}$. It is convenient
to denote such a weak value by $_{\bra{\phi}}\langle\hat{a}_{w}\rangle_{\ket{\psi}}$.''
{[}Wiseman 2007, p. 4, emphasis in the original{]}\nocite{Wiseman:2007gbmiwvab}
\end{quotation}
As the quoted passage indicates, Wiseman even went so far as to use
bracket notation in general for weak values, as a way of mimicking
the usual notation for an expectation value. This bracket notation
for weak values was also used by Aharonov, Popescu, Rohrlich, Skrzypczyk
(2013), and by Cohen, Pollak (2018)\nocite{AharonovPopescuRohrlichSkrzypczyk:2013qcc,CohenPollak:2018dowvoqouosm}.
The identification of weak values with means or averages occurs in
other papers as well, including in a review article by Dressel et
al. (2014)\nocite{DresselMalikMiattoJordanBoyd:2014cuqwvbaa}, which
calls them ``conditioned averages.''

However, the expectation value $\expectval A$ of a single-system
observable $A$ does not get its interpretation as a mean value for
a single-system observable from the AAV experimental procedure of
extracting weak values, as laid out in Subsection~\ref{subsec:The-AAV-Experimental-Protocol}.
Instead, $\expectval A$ gets its fundamental interpretation as a
mean value over the possible values $a$ in the spectrum $\spectrum\left(A\right)$
of the single-system observable $A$ from the basic definition \eqref{eq:DvNDefExpectationValue},
\begin{equation}
\expectval A\defeq\sum_{a}a\,p\left(a\right),\label{eq:DefExpectationValue}
\end{equation}
 which is manifestly a probability-weighted average of the possible
values $a$ of the single-system observable $A$, where $p\left(a\right)$
is the Born-rule probability for the measurement outcome $a$.

It is true that if an expectation value like $\expectval A$ is, practically
speaking, \emph{extracted} via the AAV experimental protocol, then
that protocol will involve averaging over the distribution \eqref{eq:WeakValuesDevProbabilityDistributionMomentumSpace}
of the measuring device's pointer variable and averaging over the
distribution \eqref{eq:WeakValuesDevProbabilityDistributionCoordSpace}
of the measuring device's conjugate coordinate. Nevertheless, this
merely \emph{practical} \emph{technique} of averaging over an ensemble
of measuring-device variables is conceptually different from the \emph{fundamental
definition} of $\expectval A$ as an average or mean over the \emph{single-system
observable} $A$. Moreover, for a weak value $A_{w}\left(\Psi^{\prime},\Psi\right)$
for which $\ket{\Psi^{\prime}}\not\propto\ket{\Psi}$, this average
or mean over an ensemble of \emph{measuring\textendash device variables}
is the \emph{only} sense in which the weak value is a mean value,
at least according to the DvN axioms. These subtly distinct senses
of the word \textquoteleft mean\textquoteright{} are yet another reason
why there have been so many debates over the proper interpretation
of weak values. 

To make clear why these conceptually different notions of \textquoteleft mean\textquoteright{}
really matter, and are not merely a form of hairsplitting, it might
help to consider a more familiar-looking example. Suppose that a teacher
running a classroom of some fixed number of students decides to assign
each student a distinct natural number, chosen at random. The teacher
then writes these numbers down on a sheet of paper, but does not show
the list to the students. Later on, the teacher could certainly \textquoteleft observe\textquoteright{}
or \textquoteleft measure\textquoteright{} this list of arbitrary
natural numbers and compute their statistical mean, but the answer
would be a mean only in the narrow sense of being an average over
measurement results, without being an average over single-student
attributes possessed by the students themselves. As such, this statistical
mean would reveal nothing whatsoever about the students. By contrast,
if the teacher instead decides to measure each student's height, or
decides to ask the students for their ages, then the teacher could
compute statistical means of a qualitatively different conceptual
character, because a mean of heights or a mean of ages would not merely
be an average of measurement results, but would be an average over
single-student attributes. In particular, a classroom visitor looking
at the statistical mean of the list of arbitrary natural numbers would
learn nothing about the students, but the visitor could learn something
interesting from the mean of heights or from the mean of ages.

The AACV paper focused on instrumentalist questions of measurement,
and said very little about the interpretation of weak values beyond
that. The AAV paper likewise described weak values mostly as the possible
outcomes of a particular class of experimental protocols.

In a 1990 paper by Aharonov and Vaidman titled ``Properties of a
Quantum System During the Time Interval Between Two Measurements''
(Aharonov, Vaidman 1990)\nocite{AharonovVaidman:1990poaqsdttibtm},
one began to see stronger claims of interpretation. The paper's title
itself suggested that a generic weak value should be understood as
a property of a single quantum system. Indeed, the paper's third paragraph
included the following statements:
\begin{quotation}
The most important outcome of our approach is the possibility to define
a new concept: the \emph{weak value} of a quantum variable. It is
a physical property of a quantum system between two measurements,
i.e., a property of a system belonging to an ensemble that is both
preselected and postselected. This property can manifest itself through
a measurement that fulfills certain requirements of weakness. {[}Ibid.,
p. 11, emphasis in the original{]}\nocite{AharonovVaidman:1990poaqsdttibtm}
\end{quotation}
Note the explicit inference of a single-system property merely from
the fact that the system belongs to a specially chosen ensemble defined
through post-selection. Again, the attempt to identify a single-system
property or observable with an ensemble observable, or to infer the
former from the latter, is precisely an example of the ensemble fallacy
\eqref{eq:DefEnsembleFallacy}. The attempt to make direct inferences
about systems through the deliberate use of post-selection implies
committing the post-selection fallacy \eqref{eq:DefPostSelectionFallacy}.

The paper tried to get around these obstructions by considering the
extraction of weak values in a \emph{single} measurement, but on a
large composite system consisting of $N\gg1$ individual spin-1/2
subsystems\textemdash which is, of course, an ensemble from the perspective
of each single spin-1/2 system. Even then, the paper concluded that
the measurement outcome could have been described as the result of
measuring a traditional observable of the large composite system according
to the standard DvN formalism: ``We shall now show how the above
result can be explained using the standard formalism, which we in
no way dispute.'' (Aharonov, Vaidman 1990, p. 16)\nocite{AharonovVaidman:1990poaqsdttibtm}

In a 1991 paper, Aharonov and Vaidman showed that for certain choices
of pre-selected state vector $\ket{\Psi}$, observable $A$, and post-selected
state vector $\ket{\Psi^{\prime}}$, the weak value $A_{w}$ coincided
with one of the observable's eigenvalues (Aharonov, Vaidman 1991)\nocite{AharonovVaidman:1991cdoaqsaagt}.
This result did not, however, establish that weak values were single-system
observables. Indeed, given certain choices of quantum state, an expectation
value $\expectval A$ can sometimes be equal to one of the eigenvalues
of the corresponding observable $A$, but that does not make expectation
values single-system observables. A quantity that is only \emph{contingently}
related to a single-system observable, in a state-dependent way, is
still categorically different from a single-system observable.

In a 1996 paper, Vaidman, inspired by the definition of an ``element
of reality'' from the famous EPR paper (Einstein, Podolsky, Rosen
1935)\nocite{EinsteinPodolskyRosen:1935cqmdprbcc}, introduced a new
definition, intended to capture weak values:
\begin{quotation}
I suggest to take this property to be the definition: \emph{If we
are certain that a procedure for measuring a certain variable will
lead to a definite shift of the unchanged probability distribution
of the pointer, then there is an element of reality: the variable
equal to this shift.} {[}Vaidman 1996a, p. 898, emphasis in the original{]}\nocite{Vaidman:1996wmeor}
\end{quotation}
It is certainly true that the full AAV experimental protocol for extracting
a weak value leads to shifts in quantum states of measuring devices,
as \eqref{eq:WeakValuesFinalStateVectorDevice} makes clear. Moreover,
the question over whether there are elements of reality distinct from
quantum states is an important topic in quantum foundations and in
debates over the interpretation of quantum theory. However, this definition
of an element of reality does not distinguish between single-system
elements of reality and ensemble elements of reality, nor does it
address the way that post-selection by an external observer is necessary
to produce the desired weak value, because, again, as emphasized in
the discussion surrounding \eqref{eq:WeakValuesTotalStateVectorAfterObsMeasurement},
the interactions between the measuring devices and the subject systems
alone are insufficient. Those issues are key to the critique of the
research literature on weak values laid out in the present work. Indeed,
as Vaidman added: 
\begin{quotation}
In such a case, a measurement performed on a single system does not
yield the value of the shift (the element of reality), but such measurements
performed on a large enough ensemble of identical systems yield the
shift with any desirable precision. {[}Ibid., p. 898{]}\nocite{Vaidman:1996wmeor}
\end{quotation}
In the conclusion of the paper, Vaidman went on to acknowledge that
weak values were not generically single-system observables, but nonetheless
made the case for taking them seriously: 
\begin{quotation}
I certainly see a deficiency of weak-measurement elements of reality
defined above in the situations in which they cannot be measured on
a single system. Still, I do not think that the fact that a weak value
cannot be measured on a single system prevents it from being a ``reality.''
We know that the measuring device shifts its pointer exactly according
to the weak value, even though we cannot find it because of the large
uncertainty of the pointer position. We can verify this knowledge
performing measurements on an ensemble. {[}Ibid., p. 904{]}\nocite{Vaidman:1996wmeor}
\end{quotation}

Weak values have been measured in many experiments in the years since
the AACV and AAV paper were published. In that regard, the AACV and
AAV papers represented a marvelous triumph. Nevertheless, without
committing the measurementist fallacy \eqref{eq:DefMeasurementistFallacy},
there is no reason to think that the success of those experiments
in obtaining matrix elements \eqref{eq:DefWeakValueGeneral} provides
any support for the interpretation of weak values presented in the
subsequent research literature as single-system observables, or as
providing a way to infer single-system observables.

\section{Questionable Applications of Weak Values\label{sec:Questionable-Applications-of-Weak-Values}}

\subsection{Quantum Cheshire Cats\label{subsec:Quantum-Cheshire-Cats}}

In 2013, Aharonov, Popescu, Rohrlich, and Skrzypczyk (APRS) published
a paper titled ``Quantum Cheshire Cats'' in the \emph{New Journal of Physics}
(Aharonov, Popescu, Rohrlich, Skrzypczyk 2013)\nocite{AharonovPopescuRohrlichSkrzypczyk:2013qcc}.
The APRS paper claimed that the location of a \emph{single photon}
could seemingly be separated from the photon's polarization. Taking
a cue from Lewis Carroll's \emph{Alice in Wonderland}, whose Cheshire
Cat could be separated from its grin, the paper called the photon
a ``quantum Cheshire Cat.''

The grounds for this claim were counterfactual reasoning with post-selection,
together with the invocation of weak values. Although the APRS paper
acknowledged that these notions all applied at the level of an ensemble
of many particles, either measured individually or all at once, the
paper nonetheless committed the ensemble fallacy \eqref{eq:DefEnsembleFallacy}
by explicitly claiming to make single-particle inferences from these
ensemble observables. Indeed, in the paper's introduction, one found
these statements:
\begin{quotation}
Yet, as we will show here, in the curious way of quantum mechanics,
photon polarization may exist where there is no photon at all. At
least this is the story that quantum mechanics tells via measurements
on a pre- and post-selected ensemble. {[}Ibid., p. 2{]}\nocite{AharonovPopescuRohrlichSkrzypczyk:2013qcc}
\end{quotation}
The APRS paper went on to add:
\begin{quotation}
Let us first ask which way the photon went inside the interferometer.
We will show that, given the pre- and post-selection, \emph{with certainty
the photon went through the left arm}. {[}...{]} Thus the non-demolition
measurement in the right arm never finds the photon there, indicating
that the photon must have gone through the left arm. {[}...{]} \emph{The
Cat is therefore in the left arm. But can we find its grin elsewhere?}
{[}...{]} We will discover that \emph{there is angular momentum in
the right arm}. {[}...{]} We seem to see what Alice saw\textemdash a
grin without a cat! We know with certainty that the photon went through
the left arm, yet we find angular momentum in the right arm. {[}Ibid.,
pp. 3\textendash 4, emphasis in the original{]}\nocite{AharonovPopescuRohrlichSkrzypczyk:2013qcc}
\end{quotation}
The use of language like ``the photon went through the left arm''
and ``The Cat is therefore in the left arm'' attributes \emph{happening}
meanings to properties of quantum systems that are distinct from measurement
statements, and therefore runs directly into the category problem
described in Subsection~\ref{subsec:The-Dirac-von-Neumann-Axioms}.
Moreover, the statements in this quoted passage are all single-particle
statements, and are therefore inadmissible due to committing the ensemble
fallacy \eqref{eq:DefEnsembleFallacy}.

The APRS paper did not address these issues, but instead suggested
a different problem:
\begin{quotation}
But could this conclusion really be right? It is, ultimately, open
to the following criticism. We never actually simultaneously measured
the location and the angular momentum. Indeed, our conclusions above
were reached by measuring location on some photons and angular momentum
on others. The immediate implication is that all we have here is a
paradox of counterfactual reasoning {[}...{]}. That is, we have made
statements about where the photon is, and about where the angular
momentum is, that are paradoxical as long as we don\textquoteright t
actually perform all the relevant measurements simultaneously. {[}Ibid.,
p. 4, emphasis in the original{]}\nocite{AharonovPopescuRohrlichSkrzypczyk:2013qcc}
\end{quotation}
The APRS paper claimed to solve this problem by appealing instead
to weak values of projectors having the special values $1$ or $0$,
but, as explained in the present work, weak values do not provide
an escape from the ensemble fallacy \eqref{eq:DefEnsembleFallacy}
or from the post-selection fallacy \eqref{eq:DefPostSelectionFallacy}.
Furthermore, as explained in Subsection~\ref{subsec:Weak-Values-for-Projectors},
the values $1$ and $0$ do not have the intuitive meanings of \textquoteleft true\textquoteright{}
and \textquoteleft false\textquoteright{} for the weak values of projectors.

Other papers since 2013 have commented on quantum Cheshire Cats both
theoretically and experimentally (for example, Denkmayr et al. 2014,
Corrêa et al. 2015, Duprey et al. 2018)\nocite{DenkmayrGeppertSponarLemmelMatzkinTollaksenHasegawa:2014ooaqcciamwie,CorreaFrancaSantosMonkenSaldanha:2015qccasqi,DupreyKanjilalSinhaHomeMatzkin:2018tqccetbaoi},
but none have approached the issue according to the specific form
of critical analysis carried out in the present work.

\subsection{Bohmian Trajectories\label{subsec:Bohmian-Trajectories}}

Weak values have also shown up in the research literature on Bohmian
mechanics (de Broglie 1930; Bohm 1952a, 1952b)\nocite{deBroglie:1930iswm,Bohm:1952siqtthvi,Bohm:1952siqtthvii},
a hidden-variables formulation or interpretation of quantum theory
in which the Hilbert-space ingredients of textbook quantum theory
are augmented with classical-like particles that have definite positions
at all times and follow hidden trajectories that are guided or piloted
by their overall wave function.

In 2007, Wiseman published a paper titled ``Grounding Bohmian Mechanics
in Weak Values and Bayesianism'' (Wiseman 2007)\nocite{Wiseman:2007gbmiwvab}
in which he argued that weak values could be used to assign an objective
physical reality to those hidden trajectories by giving an operational
definition to a specific choice of vector field: 
\begin{quotation}
\noindent Since velocity is defined by the rate of change of position,
in fact we want simply to make a weak measurement of initial position,
then a strong measurement of position a short time $\tau$ later.
We then have the following \emph{operational definition} for the velocity
for a particle at position $\mathbf{x}$: 
\begin{equation}
\mathbf{v}\left(\mathbf{x};t\right)\defeq\lim_{\tau\to0}\tau^{-1}E\left[\mathbf{x}_{\textrm{strong}}\left(t+\tau\right)-\mathbf{x}_{\textrm{weak}}\left(t\right)\given\mathbf{x}_{\textrm{strong}}\left(t+\tau\right)=\mathbf{x}\right].\qquad\left[\textrm{Wiseman's eq. }\left(5\right)\right]\label{eq:WisemanDefVelocityField}
\end{equation}
 Here $E\left[a\given F\right]$ denotes the average of $a$ over
the (post-selected) ensemble where $F$ is true. {[}Ibid., p. 4, emphasis
in the original{]}\nocite{Wiseman:2007gbmiwvab}
\end{quotation}
It is true that the ensemble-level vector field $\mathbf{v}\left(\mathbf{x};t\right)$,
as defined above, has measurement units of a velocity, but Wiseman
explicitly attributes it to ``a particle at position $\mathbf{x}$,''
thereby triggering the ensemble fallacy \eqref{eq:DefEnsembleFallacy}.
Moreover, the weak value $\mathbf{x}_{\textrm{weak}}\left(t\right)$
implicitly depends on the post-selection condition $\mathbf{x}_{\textrm{strong}}\left(t+\tau\right)=\mathbf{x}$,
so to make direct inferences about the particle based on this post-selection
condition would mean committing the post-selection fallacy \eqref{eq:DefPostSelectionFallacy}.
Finally, the weak value $\mathbf{x}_{\textrm{weak}}\left(t\right)$
itself does not have a clear physical interpretation, as the present
work has argued, so the difference between $\mathbf{x}_{\textrm{strong}}\left(t+\tau\right)-\mathbf{x}_{\textrm{weak}}\left(t\right)$
does not have a clear physical meaning, either.

An interesting feature of the vector field $\mathbf{v}\left(\mathbf{x};t\right)$
defined in \eqref{eq:WisemanDefVelocityField} is that, as Wiseman
explains, it coincides with the definition of the velocity of a Bohmian
particle located at position $\mathbf{x}$ and time $t$ for a suitable
choice of Hamiltonian. However, if this observation is used to justify
the definition \eqref{eq:WisemanDefVelocityField}, then \eqref{eq:WisemanDefVelocityField}
cannot then be used to justify the physical reality of Bohmian trajectories,
on pain of logical circularity.

To avoid this logical circularity, Wiseman attempts to provide an
independent justification for \eqref{eq:WisemanDefVelocityField}: 
\begin{quotation}
Consider a naive experimentalist, with no knowledge of QM beyond the
following basic facts about experiments at the microscopic scale:
(i) no matter how carefully a preparation procedure is repeated, the
measured properties of the particle will vary in different runs of
the experiment; (ii) if a given property of the particle is measured
strongly (arbitrarily accurately) then in general this will drastically
alter the future distribution of measurement results; (iii) if an
arbitrarily weak measurement is used instead, the future distribution
can remain essentially unaltered. For such an experimentalist, equation
(5) would be the only sensible way to measure the velocity of a particle
at position $\mathbf{x}$. Thus I will call the velocity in equation
(5) the \emph{naively observable velocity}, and contend that it is
the most natural operational definition of velocity. {[}Ibid., p.
5, emphasis in the original{]}\nocite{Wiseman:2007gbmiwvab}
\end{quotation}
The first two desiderata (i) and (ii) are certainly facts about presently-known
kinds of experiments at the microscopic scale, and are fully in keeping
with the DvN axioms, as laid out in Subsection~\ref{subsec:The-Dirac-von-Neumann-Axioms}.

The third desideratum (iii), however, is not a fact, because a fact
cannot rely fundamentally on an ill-defined concept. Notice that (iii)
depends on the notion of a ``weak measurement,'' as described in
Subsection~\ref{subsec:The-AAV-Experimental-Protocol}, together
with the implicit assumption that the weak value extracted from a
weak measurement has some sort of clear physical meaning. Without
being able to attach a clear physical meaning to a weak value, especially
in light of the dangers of post-selection that are involved in constructing
one, (iii) does not offer any means of obtaining any other physical
notions, let alone the notion of a single-particle velocity. To argue
otherwise by appealing to the fact that the vector field $\mathbf{v}\left(\mathbf{x};t\right)$
consists of quantities that can be obtained experimentally, and that
indeed \emph{have} been obtained in later experimental work (Kocsis
et al. 2011; Mahler et al. 2016)\nocite{KocsisBravermanRavetsStevensMirinShalmSteinberg:2011otatospiatsi,MahlerRozemaFisherVermeydenReschWisemanSteinberg:2016enasbt},
would be to commit the measurementist fallacy \eqref{eq:DefMeasurementistFallacy}.
(For further arguments against the interpretation of $\mathbf{v}\left(\mathbf{x};t\right)$
as describing single-particle velocities, see Fankhauser, Dürr 2021.)\nocite{FankhauserDurr:2021hntuwmov}

\section{Conclusion\label{sec:Conclusion}}

The present work provided a critical analysis of interpretational
claims made about weak values, arguing that while weak values are
clearly useful for practical applications, like signal amplification
and state tomography, weak values do not reveal interesting physical
or metaphysical properties of individual quantum systems. Since the
first appearance of weak values in the research literature, there
were great hopes that they could provide a new window into the inner
workings of quantum theory. Unfortunately, nature has not been as
forthcoming with its secrets as one might have wished.

As in other work (Barandes 2026)\nocite{Barandes:2026taratpops}, 
a larger issue is the use of post-selection in quantum-foundations
papers. As statisticians know, post-selection can lead to spurious
correlations and other statistical artifacts. Distinguishing those
sorts of statistical artifacts from truly quantum behavior can be
challenging. As a general rule, then, papers that deliberately deploy
post-selection in the course of deriving exotic results should be
especially careful to explain why those results are really due to
quantum theory itself. Research journals should be sure to hold authors
to that standard.

\section*{Acknowledgments}

The author would like to thank David Albert, David Kagan, Logan McCarty,
Xiao-Li Meng, Filip Niewinski, John Norton, James Robins, Yossi Sirote,
Kelly Werker Smith, Lev Vaidman, Larry Wasserman, Michael Weissman,
and Ken Wharton for helpful discussions.

\bibliographystyle{1_home_jacob_Documents_Work_My_Papers_2025-The_ABL_Rule_and_Weak_Values_custom-abbrvalphaurl}
\bibliography{0_home_jacob_Documents_Work_My_Papers_Bibliography_Global-Bibliography}

\end{document}